\documentclass[aps,twocolumn]{revtex4}

\RequirePackage[T2A]{fontenc}

\usepackage{amssymb}
\usepackage{amsmath}
\usepackage{bm}
\usepackage[dvips]{epsfig}
\usepackage{graphicx}

\begin{document}

\title{Lifshitz transitions and angular conductivity diagrams 
in metals with complex Fermi surfaces
}

\author{A.Ya. Maltsev}

\affiliation{
\centerline{\it V.A. Steklov Mathematical Institute of Russian Academy of Sciences}
\centerline{\it 119991 Moscow, Gubkina str. 8}
}

\begin{abstract}
We consider the Lifshitz topological transitions and the corresponding 
changes in the galvanomagnetic properties of a metal from the point of 
view of the general classification of open electron trajectories arising 
on Fermi surfaces of arbitrary complexity in the presence of magnetic field.
The construction of such a classification is the content of the Novikov 
problem and is based on the division of non-closed electron trajectories 
into topologically regular and chaotic trajectories. The description of 
stable topologically regular trajectories gives a basis for a complete 
classification of non-closed trajectories on arbitrary Fermi surfaces 
and is connected with special topological structures on these surfaces.
Using this description, we describe here the distinctive features of 
possible changes in the picture of electron trajectories during the Lifshitz 
transitions, as well as changes in the conductivity behavior in the presence 
of a strong magnetic field. As it turns out, the use of such an approach makes 
it possible to describe not only the changes associated with stable electron 
trajectories, but also the most general changes of the conductivity diagram 
in strong magnetic fields.
\end{abstract}

\maketitle

\section{Introduction}

 In this paper, we will try to describe the most general relationship 
between the Lifshitz transitions (see \cite{Lifshits1960,etm}), leading 
to a change in the topology of the Fermi surface, and angular diagrams 
that describe the behavior of the magnetic conductivity of a metal in 
strong magnetic fields. It must be said that at present topological 
Lifshitz transitions are actually represented by a very wide range of 
phenomena associated with topological properties of a Fermi surface 
and their changes, and the study of the variety of such phenomena is 
an interesting and rapidly developing area of condensed matter physics 
(see, for example, \cite{Volovik2017,Volovik2018}).

\begin{figure}[t]
\begin{center}
\includegraphics[width=0.9\linewidth]{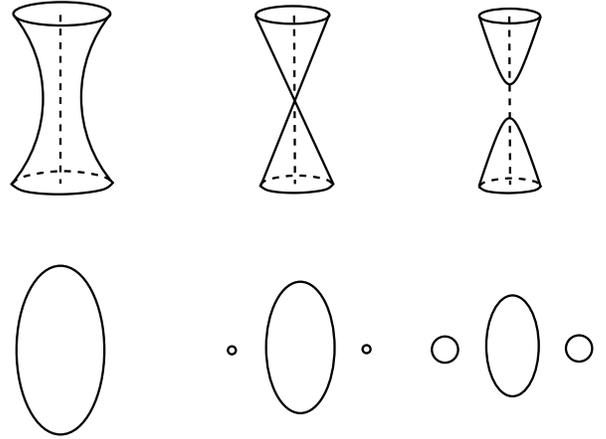}
\end{center}
\caption{Reconstruction of the Fermi surface and arising
of new components when passing through the critical points of the 
relation $\epsilon ({\bf p})$ (\cite{Lifshits1960})
}
\label{Fig1}
\end{figure}

 Here, however, we will consider the most classical definition of 
the Lifshitz transitions (\cite{Lifshits1960}), namely, a change in the 
topology of the Fermi surface when passing the critical points of the 
dispersion relation $\epsilon ({\bf p})$ (see Fig. \ref{Fig1}).

 As is well known, the dispersion relation $\epsilon ({\bf p})$ 
can be considered either as a periodic function in the quasi-momentum 
space $\mathbb{R}^{3}$, or as a smooth function on the three-dimensional 
torus $\mathbb{T }^{3}$ obtained from $\mathbb{R}^{3}$ by factorization 
with respect to the reciprocal lattice vectors. The singular points of 
the function $\epsilon ({\bf p})$ are defined by the condition 
$\, \nabla \epsilon ({\bf p}) = 0 \, $, and the corresponding energy levels, 
as is known, correspond to arising of the Van Hove singularities in the 
density of electron states.

 The singularities of the function $\epsilon ({\bf p})$ include the 
points of its local minima and maxima, as well as saddle singular points 
(assuming that all singular points of $\epsilon ({\bf p})$ are non-degenerate).

 Saddle points of a function in three-dimensional space, as is well known, 
can have index 1 or 2, depending on whether the increment of the function 
near this point can be represented in the form
$$ d \epsilon ({\bf p}) \, = \, a^{2} d p_{1}^{2} + b^{2} d p_{2}^{2}
- c^{2} d p_{3}^{2} $$
or
$$ d \epsilon ({\bf p}) \, = \, a^{2} d p_{1}^{2} - b^{2} d p_{2}^{2}
- c^{2} d p_{3}^{2} $$
in some local Euclidean coordinate system. As is well known from the Morse 
theory, the number of saddle singular points of both types for a function on a 
three-dimensional torus is always at least three. In fact, for real dispersion 
laws, it is often larger, in particular, whenever Fermi surfaces of genus 
greater than 3 arise. Here, we are interested precisely in the saddle 
singularities of the relation $\epsilon ({\bf p})$.

 As was shown in \cite{Lifshits1960}, the passage of the Fermi level 
through critical points of $\epsilon ({\bf p})$ 
(for example, under strong external pressure) leads to singularities 
in the thermodynamic quantities of the electron gas in the crystal 
(the Lifshitz transitions), as well as possible abrupt changes in the 
behavior of the magnetic conductivity in strong magnetic fields. The latter 
circumstance is associated with a possible significant 
change in the geometry of the trajectories of the system
\begin{equation}
\label{MFSyst}
{\dot {\bf p}} \,\,\,\, = \,\,\,\, {e \over c} \,\,
\left[ {\bf v}_{\rm gr} ({\bf p}) \, \times \, {\bf B} \right]
\,\,\,\, = \,\,\,\, {e \over c} \,\, \left[ \nabla \epsilon ({\bf p})
\, \times \, {\bf B} \right] \,\,\, ,
\end{equation}
describing the semiclassical dynamics of electrons in an external magnetic 
field, on the Fermi surface. The main effect here is a sharp change in the 
behavior of the magnetic conductivity due to arising (or disappearance) 
of open trajectories of system (\ref{MFSyst}) during topological 
reconstructions of the Fermi surface.

 The important role of open trajectories of system (\ref{MFSyst}) 
in description of the conductivity of metals in strong magnetic 
fields was also first revealed by the school of I.M. Lifshitz 
(see \cite{lifazkag,lifpes1,lifpes2,etm}). Since the trajectories of 
system (\ref{MFSyst}) are defined by the intersections of the 
surfaces $\, \epsilon ({\bf p}) = {\rm const} \, $ by planes orthogonal 
to the magnetic field, the geometry of such trajectories is essentially 
determined by the geometry and the topology of the Fermi surface.
In particular, the question of whether the Fermi surface is bounded 
or unbounded in the ${\bf p}$ - space is of great importance.

 As was shown in \cite{lifazkag}, the contributions of closed and open 
periodic trajectories to the conductivity tensor differ significantly 
in the limit $\, \omega_{B} \tau \rightarrow \infty \, $ 
(i.e., in the limit of strong magnetic fields). In particular, if there 
are only closed trajectories on the Fermi surface, the conductivity 
decreases in all directions in the plane orthogonal to ${\bf B}$ 
in the specified limit. The asymptotic behavior of the total 
conductivity tensor can then be represented in the form
\begin{equation}
\label{Closed}
\sigma^{kl} \,\,\,\, \simeq \,\,\,\,
{n e^{2} \tau \over m^{*}} \, \left(
\begin{array}{ccc}
( \omega_{B} \tau )^{-2}  &  ( \omega_{B} \tau )^{-1}  &
( \omega_{B} \tau )^{-1}  \cr
( \omega_{B} \tau )^{-1}  &  ( \omega_{B} \tau )^{-2}  &
( \omega_{B} \tau )^{-1}  \cr
( \omega_{B} \tau )^{-1}  &  ( \omega_{B} \tau )^{-1}  &  *
\end{array}  \right)  \quad  ,  
\end{equation}
($\omega_{B} \tau \, \rightarrow \, \infty$).

 At the same time, the contribution of open periodic trajectories to 
the conductivity tensor is strongly anisotropic in the plane orthogonal 
to ${\bf B}$ and can be represented in the leading order as
\begin{equation}
\label{Periodic}
\sigma^{kl} \,\,\,\, \simeq \,\,\,\,
{n e^{2} \tau \over m^{*}} \, \left(
\begin{array}{ccc}
( \omega_{B} \tau )^{-2}  &  ( \omega_{B} \tau )^{-1}  &
( \omega_{B} \tau )^{-1}  \cr
( \omega_{B} \tau )^{-1}  &  *  &  *  \cr
( \omega_{B} \tau )^{-1}  &  *  &  *
\end{array}  \right)  \quad  , 
\end{equation}
($\omega_{B} \tau \, \rightarrow \, \infty$). 

 In the formulas (\ref{Closed}) - (\ref{Periodic}), as everywhere 
further, it is assumed that the $z$ axis is directed along the magnetic 
field. In the relation (\ref{Periodic}), it is also assumed that the 
direction of the axis $x$ coincides with the mean direction of the open 
trajectories in the ${\bf p}$ -space. The sign $\simeq$ in both formulas 
means asymptotic behavior, i.e. each of the components actually contains 
some dimensionless factor of order 1. The quantity $\omega_{B}$ plays the 
role of the electron cyclotron frequency in the metal, while the quantity 
$\tau$ represents the mean free time of electrons. The quantity $m^{*}$ 
has the meaning of the effective mass of an electron in a crystal.
The relation $\, \omega_{B} \tau \gg 1 \, $, as is also well known, 
requires the use of sufficiently pure single-crystal samples at very 
low temperatures ($T \leq 1 K$) and sufficiently strong magnetic fields 
($B \geq 1 Tl$).

 The value $n$ usually plays the role of the concentration of current 
carriers in the metal. In the formulas (\ref{Closed}) - (\ref{Periodic}), 
however, it is also proportional to the measure of the corresponding 
trajectories (closed or periodic) on the Fermi surface. The latter 
circumstance is especially important in the situation we are considering, 
since the measure of open trajectories can be determined by the proximity 
to the Lifshitz transition point $\epsilon_{0}$. It is this situation 
that occurs, for example, in \cite{Lifshits1960}, where the arising
and disappearance of periodic open trajectories on a ``warped cylinder'' 
surface is considered. In this situation, the leading term of the 
conductivity tensor in the presence of open trajectories on the Fermi 
surface was represented in \cite{Lifshits1960} in the form 
\begin{equation}
\label{FullContr}
\sigma^{kl}  =  
\left(
\begin{array}{ccc}
\gamma^{2} a_{xx} &  \gamma a_{xy} &  \gamma a_{xz} \cr
\gamma a_{yx} & \gamma^{2} a_{yy} + \beta^{1/2} b_{yy} &
\gamma a_{yz} + \beta^{1/2} b_{yz} \cr
\gamma a_{zx} & \gamma a_{zy} + \beta^{1/2} b_{zy} & a_{zz} 
\end{array}  \right) 
\end{equation}
($\omega_{B} \tau \, \rightarrow \, \infty$),
where $\, \gamma = (\omega_{B} \tau )^{-1} \, $, 
$\, \beta = |(\epsilon_{F} - \epsilon_{0}) / \epsilon_{F}| \, $,
the quantities $a_{kl}$ represent some constants, and the quantities 
$b_{kl}$ have a weak (logarithmic) dependence on $\beta$.

 As we can see, the formula (\ref{FullContr}) allows not only to 
observe the Lifshitz transition in the described situation, but also 
to determine the proximity to this transition when changing the 
parameters of influence on a sample.

\vspace{1mm}
 
\begin{figure}[t]
\begin{center}
\includegraphics[width=\linewidth]{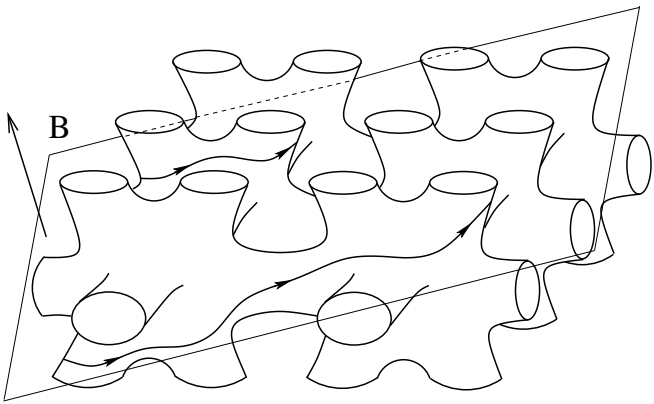}
\end{center}
\caption{Trajectories of system (\ref{MFSyst}) on a general periodic 
Fermi surface}
\label{Fig2}
\end{figure}

 In the general situation, the Fermi surface is an arbitrary 3-periodic 
surface in ${\bf p}$ - space (Fig. \ref{Fig2}), and the problem of 
describing the trajectories of the system (\ref{MFSyst}) is quite difficult.
For the first time, the problem of complete classification of open 
trajectories of system (\ref{MFSyst}) was set by S.P. Novikov 
in \cite{MultValAnMorseTheory} and then intensively studied in his 
topological school 
(see \cite{zorich1,dynn1992,Tsarev,dynn1,zorich2,DynnBuDA,dynn2,dynn3}).
As a result of studying the Novikov problem, a number of deep topological 
results have been obtained, and by now a complete classification of the 
open trajectories of system (\ref{MFSyst}) for arbitrary periodic 
dispersion relations $\epsilon ({\bf p})$ has been obtained. 
Consequences from the topological theorems obtained in the study of 
the Novikov problem also led to a description of a number of physical 
effects associated with the behavior of open trajectories of (\ref{MFSyst}) 
and, in addition, made it possible to give a complete classification of 
possible asymptotic behavior of conductivity in strong magnetic fields 
for metals with arbitrarily complex Fermi surfaces (see e.g. \cite{PismaZhETF,ZhETF1997,UFN,JournStatPhys,TrMian,ObzorJetp}).

 Here we are interested in changes in the open trajectories of 
system (\ref{MFSyst}) under the Lifshitz transitions, i.e. changes 
in the topology of the Fermi surface when passing singular points of 
the dispersion relation $\epsilon ({\bf p})$. We will assume here 
that the Fermi surface has the most general form, and in describing 
the trajectories we will use the general classification obtained in 
the study of the Novikov problem. To describe the situation, we will 
use angular diagrams that specify the type of trajectories of system 
(\ref{MFSyst}) on the Fermi surface depending on the direction of the 
magnetic field. The angular diagram is thus the unit sphere 
$\mathbb{S}^{2}$, which parametrizes the directions of ${\bf B}$ 
and determines the type of trajectories on the Fermi surface for 
each direction. Since the type of trajectories of system 
(\ref{MFSyst}) determines the asymptotic behavior of the conductivity 
tensor in the limit of strong magnetic fields, it is natural to also 
call such diagrams conductivity (magnetic conductivity) diagrams 
for a given Fermi surface. We are mainly interested here in the 
changes in such diagrams that accompany the Lifshitz transitions. 
Experimental observation of changes (sharp jumps) in such 
diagrams can generally serve as one of the tools for studying 
the Lifshitz transitions in metals with complex Fermi surfaces.

 Here we present the general picture of changes in the conductivity 
diagrams when passing the singularities of the relation 
$\epsilon ({\bf p})$, based on the general theory of such diagrams, 
constructed in the study of the Novikov problem. In the next section,
we present the general classification of the diagrams corresponding 
to various Fermi surfaces and describe their connection with 
the angular diagrams for the entire dispersion relation
$\epsilon ({\bf p})$. In section 3, we will describe typical changes 
in angular diagrams corresponding to topological transitions of 
various types on Fermi surfaces of arbitrary complexity.

\section{General Facts about Angular Conductivity Diagrams in Metals}
\setcounter{equation}{0}

 The basis for classifying the open trajectories of system (\ref{MFSyst}) 
is the description of its stable open trajectories. Here we call open 
trajectories of (\ref{MFSyst}) stable if they do not vanish and retain 
their global geometry under small variations of all problem parameters, 
in particular, small variations of the level $\epsilon_{F}$ and rotations 
of the direction of ${\bf B }$. As follows from the results of 
\cite{zorich1,dynn1992,dynn1}, stable open trajectories of system 
(\ref{MFSyst}) have the following remarkable properties.

\vspace{1mm}

\begin{figure}[t]
\begin{center}
\includegraphics[width=\linewidth]{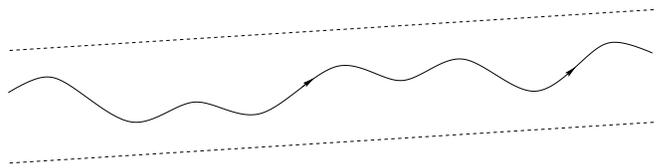}
\end{center}
\caption{The form of a stable open trajectory of system (\ref{MFSyst}) 
in a plane orthogonal to ${\bf B}$ (schematically)}
\label{Fig3}
\end{figure}

1) Each stable open trajectory of system (\ref{MFSyst}) lies in a straight 
strip of finite width in a plane orthogonal to ${\bf B}$ and passes through 
it (Fig. \ref{Fig3}).

\vspace{1mm}

2) The mean direction of all stable open trajectories  
for a given direction of the magnetic field is given by the intersection 
of the plane orthogonal to ${\bf B}$ and some integral (generated by two 
reciprocal lattice vectors) plane $\Gamma$, the direction of which is 
invariable for small variations of the problem parameters.

\vspace{1mm}

 Property (1) of stable open trajectories manifests itself directly 
in the behavior of the magnetic conductivity in strong magnetic fields. 
Namely, here, as in the case of periodic open trajectories, there is 
a strong anisotropy of the conductivity in the plane orthogonal to 
${\bf B}$, and the main term in the asymptotics of the conductivity 
tensor is also given by the formula (\ref{Periodic}). For special 
directions of ${\bf B}$, stable open trajectories (\ref{MFSyst}) can be 
periodic. In the generic case, however, they are quasi-periodic 
and have no periods in the ${\bf p}$ -space. The direction of maximum 
suppression of conductivity belongs to the corresponding plane $\Gamma$, 
which makes it experimentally observable (\cite{PismaZhETF,UFN}).
An integral plane in ${\bf p}$ - space can also be defined as a plane 
orthogonal to some integer direction of the original crystal lattice.
The plane $\Gamma$ can thus be defined by some irreducible integer 
triple $(m^{1}, m^{2}, m^{3})$. The numbers $(m^{1}, m^{2}, m^{3})$ 
were defined in \cite{PismaZhETF} as topological numbers observed in 
the conductivity of normal metals.

\vspace{1mm}

 Each family of stable open trajectories is defined by some region 
(stability zone) $\Omega$ on the angular diagram corresponding to the 
same values $(m^{1}, m^{2}, m^{3})$. In the general case, an angular 
diagram may contain some (finite or infinite) number of stability zones 
$\Omega_{\alpha}$ corresponding to different values of 
$(m^{1}_{\alpha}, m^{2}_{\alpha }, m^{3}_{\alpha})$. The presence of 
stability zones and their location on the unit sphere $\mathbb{S}^{2}$ 
is an important component of the diagram of conductivty of a metal in 
strong magnetic fields.

\vspace{1mm}

 In addition to stable open trajectories of system (\ref{MFSyst}), 
there are also unstable open trajectories. First of all, they may include 
periodic trajectories, for example, those considered above. Periodic 
trajectories are, in a sense, semi-stable, namely, they are preserved 
under rotations of ${\bf B}$ orthogonal to their mean direction, and 
collapse under all other rotations. These trajectories correspond 
to one-dimensional arcs on the angular diagram, which mark the presence 
of such trajectories on the Fermi surface for the corresponding directions 
of ${\bf B}$. The set of corresponding arcs on the sphere $\mathbb{S}^{2}$ 
is also an important part of a metal magnetic conductivity diagram.

\vspace{1mm}

 Periodic open trajectories, however, are not the only type of unstable 
open trajectories of (\ref{MFSyst}). Namely, there are open trajectories 
of system (\ref{MFSyst}) of much more complex geometry, which are unstable 
both with respect to small rotations of ${\bf B}$ and small variations of 
the Fermi level $\epsilon_{F}$ (\cite{Tsarev,DynnBuDA,dynn2}). Such 
trajectories can be conditionally divided into two main types, namely, 
Tsarev-type trajectories and Dynnikov-type trajectories. Tsarev-type 
trajectories can only be observed for partially irrational directions of 
${\bf B}$, when the plane orthogonal to ${\bf B}$ contains one 
(up to proportionality) reciprocal lattice vector. On the contrary, 
Dynnikov-type trajectories can arise only for directions of ${\bf B}$ of 
complete irrationality (the plane orthogonal to ${\bf B}$ does not contain 
reciprocal lattice vectors).

 Unstable trajectories of both types have rather complex behavior on the 
Fermi surface, which in this case should itself have sufficient complexity. 
However, the behavior of Tsarev-type trajectories in planes orthogonal 
to ${\bf B}$ is much simpler than that of Dynnikov-type trajectories.
Namely, Tsarev-type trajectories have an asymptotic direction that is 
the same in all planes orthogonal to ${\bf B}$ for a given direction of 
${\bf B}$. As a consequence, the contribution of Tsarev-type trajectories 
to the conductivity tensor also has strong anisotropy in the plane 
orthogonal to ${\bf B}$ and is close in form to the contribution 
(\ref{Periodic}), although it differs from it in some details.

\begin{figure}[t]
\begin{center}
\includegraphics[width=\linewidth]{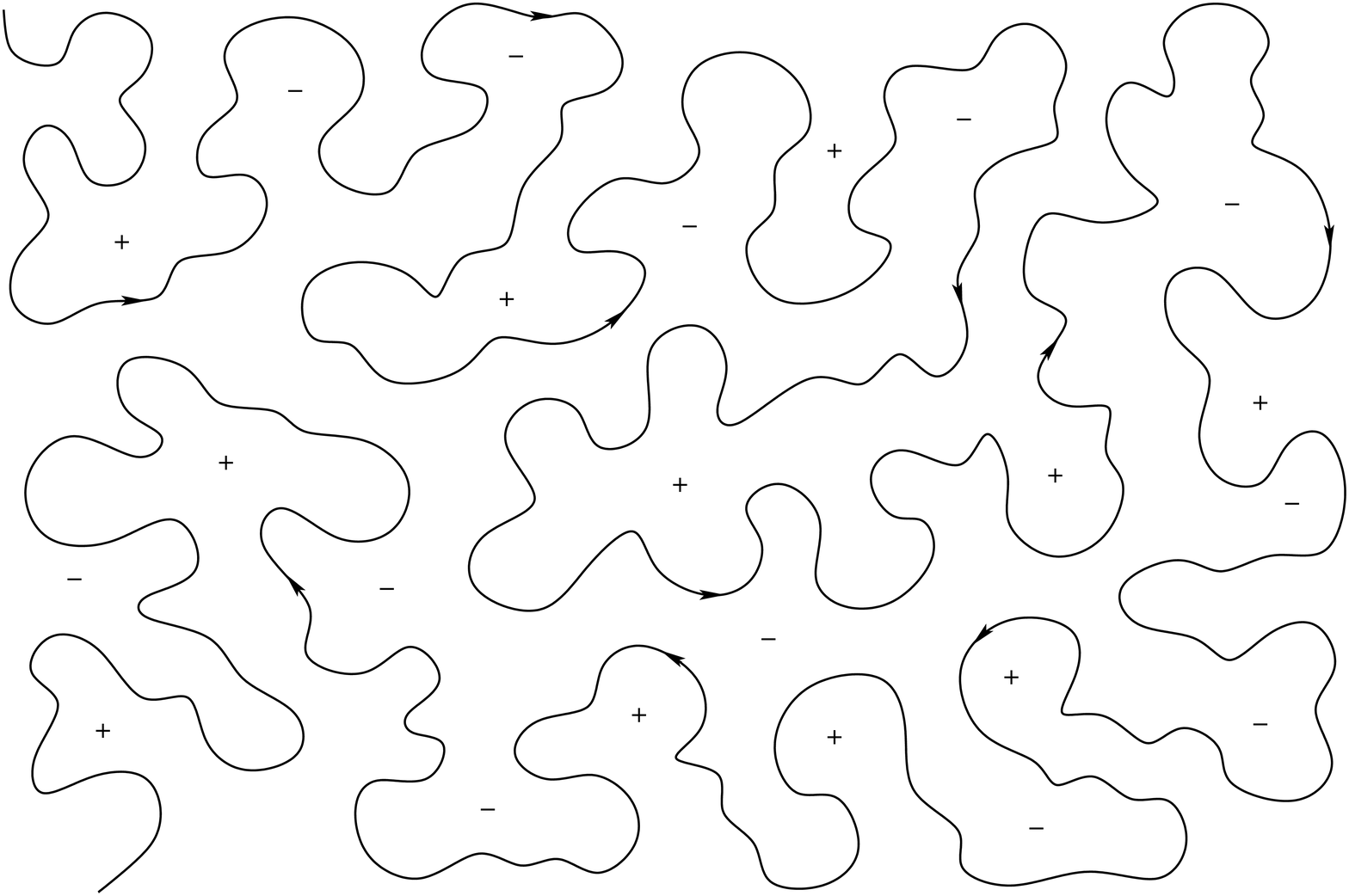}
\end{center}
\caption{Form of the Dynnikov chaotic trajectory in a plane 
orthogonal to ${\bf B}$ (schematically)
}
\label{Fig4}
\end{figure}

 Dynnikov-type trajectories have much more complex behavior in planes 
orthogonal to ${\bf B}$, wandering along them in a rather chaotic manner 
(Fig. \ref{Fig4}). Among the features of the contribution of such 
trajectories to the magnetic conductivity tensor, one can distinguish 
the suppression of conductivity along the direction of the magnetic field 
(see \cite{ZhETF1997}), as well as the arising of fractional powers of 
the parameter $\omega_{B} \tau$ in the asymptotics of the tensor components 
in the limit $\, \omega_{B} \tau \rightarrow \infty \, $ 
(\cite{ZhETF1997,TrMian}). We note here that the study of arising 
and geometric properties of Dynnikov-type trajectories is an actively 
developing area at the present time (see, for example,
\cite{Tsarev,DynnBuDA,dynn2,zorich2,Zorich1996,ZorichAMS1997,
ZhETF1997,zorich3,DeLeo1,DeLeo2,ArxivObzor,BullBrazMathSoc,
JournStatPhys,ZorichLesHouches,DeLeoDynnikov1,DeLeoDynnikov2,
Dynnikov2008,Skripchenko1,Skripchenko2,DynnSkrip1,DynnSkrip2,
AvilaHubSkrip1,AvilaHubSkrip2,TrMian,DeLeoObzor,UMNObzor}). 

 In view of the particular complexity of the geometry of trajectories 
of the Tsarev or Dynnikov type, such trajectories are usually called 
chaotic. Stable open trajectories of the system (\ref{MFSyst}), as well as 
periodic trajectories, are called topologically regular.

 The arising of unstable trajectories of the Tsarev or Dynnikov 
type on the Fermi surface is associated with particularly complex 
angular diagrams, which we describe below. The location of the 
corresponding directions of ${\bf B}$ in such diagrams is perhaps 
the most interesting information about the Fermi surface.

\vspace{1mm}

 Before describing the types of diagrams corresponding to fixed Fermi 
surfaces, it is convenient to describe the angular diagrams corresponding 
to the entire dispersion relation $\epsilon ({\bf p})$. Such diagrams were 
introduced in \cite{dynn3} and are based on an important property of open 
trajectories of (\ref{MFSyst}), namely, the type of open trajectories of 
system (\ref{MFSyst}) for a given direction of ${\bf B} $ is the same for 
all energy levels $\, \epsilon ({\bf p}) = {\rm const} \, $ at which they 
appear. Moreover, the situation in the general case can be described as 
follows (\cite{dynn3}).

\vspace{1mm}

 Consider a smooth 3-periodic function $\epsilon ({\bf p})$ whose values 
lie in the interval $[ \epsilon_{\min} , \epsilon_{\max} ]$. Consider some 
fixed direction of ${\bf B}$ and the corresponding system (\ref{MFSyst}). 
Let us assume for simplicity that the direction of ${\bf B}$ is not rational. 
Then the following assertions can be formulated.

\vspace{1mm}

\noindent
1) Open trajectories of system (\ref{MFSyst}) appear either in some 
closed energy interval
$$ \epsilon_{\min} \,\,\, < \,\,\,
\epsilon_{1} ({\bf B})  \,\,\, \leq \,\,\,
\epsilon ({\bf p}) \,\,\, \leq \,\,\,
\epsilon_{2} ({\bf B}) \,\,\, < \,\,\, \epsilon_{\max} $$
or only at one energy level
$\, \epsilon_{0} \, = \, \epsilon_{1} ({\bf B}) \, = \, 
\epsilon_{2} ({\bf B}) \, $.

\vspace{1mm}

\noindent
2) In the case
$\epsilon_{1} ({\bf B}) < \epsilon_{2} ({\bf B})$,
all nonsingular open trajectories in the interval
$\left[ \epsilon_{1} ({\bf B}) , \epsilon_{2} ({\bf B}) \right]$
lie in straight strips of finite width in planes orthogonal to 
${\bf B}$ and pass through them (Fig. \ref{Fig3}).
All of them have the same mean direction given by the intersection 
of the plane orthogonal to ${\bf B}$ with some integral plane $\Gamma$ 
in the ${\bf p}$-space.

\vspace{1mm}

\noindent
3) For generic directions of ${\bf B}$ the values
$\epsilon_{1} ({\bf B})$ and $\epsilon_{2} ({\bf B})$ coincide with 
the values of some continuous functions $ \tilde{\epsilon}_{1} ({\bf B})$ 
and $\tilde{\epsilon}_{2} ( {\bf B})$ defined everywhere on $\mathbb{S}^{ 2}$. 
However, for directions of ${\bf B}$ corresponding to arising of 
periodic open trajectories, the values of $\epsilon_{1} ({\bf B})$ and 
$\epsilon_{2} ( {\bf B})$ have ``jumps'' with the following inequalities 
$$\epsilon_{1} ({\bf B}) \, \leq \, \tilde{\epsilon}_{1} ({\bf B}) 
\, \leq \, \tilde{\epsilon}_{2} ({\bf B}) \, \leq \,
\epsilon_{2} ({\bf B}) $$

\vspace{1mm}

\noindent
4) The property
$\tilde{\epsilon}_{1} ({\bf B}) < \tilde{\epsilon}_{2} ({\bf B})$,
and the integral plane $\Gamma$ are stable with respect to small 
rotations of ${\bf B}$, so that each of the planes $\Gamma_{\alpha}$ 
corresponds to a certain ``stability zone'' $\widehat{\Omega} _{\alpha}$ 
in the space of directions of ${\bf B}$.

\vspace{1mm}

 According to \cite{dynn3}, the picture of stability zones for an arbitrary 
dispersion relation $\epsilon ({\bf p})$ can correspond to only one of the 
following situations.

\vspace{1mm}

\noindent
1) The entire unit sphere is the only stability zone $\widehat{\Omega}$ 
corresponding to some integral plane $\Gamma$.

\vspace{1mm}

\begin{figure}[t]
\begin{center}
\includegraphics[width=0.9\linewidth]{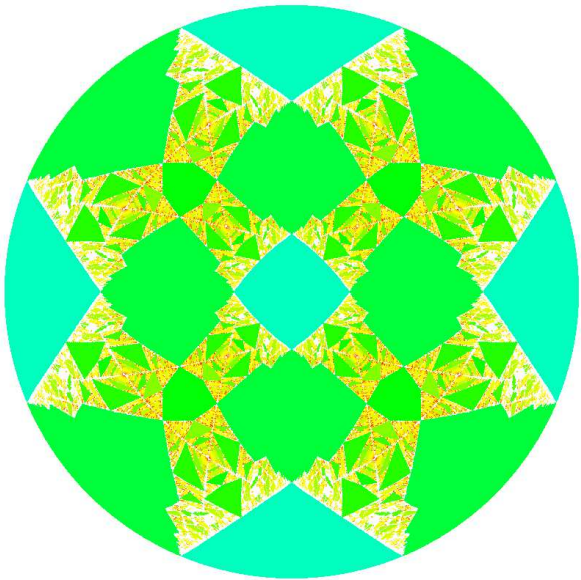}
\end{center}
\caption{Stability zones for the dispersion relation
$\epsilon ({\bf p}) = \cos p_{x} \cos p_{y} + \cos p_{y} \cos p_{z} + 
\cos p_{z} \cos p_{x}$ (\cite{DeLeoObzor})
}
\label{Fig5}
\end{figure}

\noindent
2) The angular diagram contains an infinite number of stability zones whose 
union is everywhere dense on the sphere $\mathbb{S}^{2}$ 
(see, for example, Fig. \ref{Fig5}).

\vspace{1mm}

 Case (1), as a rule, is observed for dispersion relations of a rather 
special form, close to the dispersion relations in quasi-one-dimensional 
conductors. For the vast majority of real dispersion relations, however, 
case (2) takes place. We will call here the dispersion relations corresponding 
to case (1) dispersion relations with simple angular diagram. Similarly, 
the dispersion relations corresponding to case (2) will be called relations 
with a complex angular diagram. Here we are primarily interested in dispersion 
relations with complex angular diagrams.

\vspace{1mm}

 The complement ${\cal M}$ to the union of stability zones is a rather 
complex set of fractal type on the sphere $\mathbb{S}^{2}$. According to 
the conjecture of S.P. Novikov (\cite{ArxivObzor}), this set has measure 
zero and the fractal dimension strictly less than 2. The first part of 
the Novikov conjecture was recently proved for dispersion relations 
satisfying the additional condition 
$\epsilon(-{\bf p}) = \epsilon ({\bf p})$ 
(I.A. Dynnikov, P. Hubert, P. Mercat, and A.S. Skripchenko, in the process 
of publication). The second part of the conjecture is confirmed by serious 
numerical studies, but has not yet been proved rigorously.

 The points of the set ${\cal M}$ represent accumulation points 
of decreasing stability zones. Moreover, the set ${\cal M}$ can contain 
special rational directions of ${\bf B}$ (see \cite{DynMalNovUMN}), as well as 
directions of ${\bf B}$ corresponding to arising of trajectories of the 
Tsarev or Dynnikov type. The set of special rational directions of ${\bf B}$, 
however, is only a countable subset of the set ${\cal M}$, so that ``almost all'' 
points of the set ${\bf M}$ represent, in fact, ``chaotic'' directions of these 
two types. The values of the functions $\tilde{\epsilon}_{1} ({\bf B})$ and 
$\tilde{\epsilon}_{2} ( {\bf B})$ coincide on the set ${\cal M} $, 
as well as on the boundaries of all the zones $\widehat{\Omega}_{\alpha}$. 

\vspace{1mm}

 It is easy to see that open trajectories appear on a fixed Fermi surface 
for a given direction of ${\bf B}$ only if the Fermi level falls within the 
corresponding interval $[\epsilon_{1} ({\bf B}) , \epsilon_{2} ({\bf B})]$. 
In particular, each stability zone $\Omega_{\alpha}$ at the conductivity 
diagram is a subdomain of some zone $\widehat{\Omega}_{\alpha}$ defined for 
the entire dispersion relation. As a rule, most of the conductivity diagram 
of a metal is in fact the region corresponding to the presence of only 
closed trajectories on the Fermi surface.

\begin{figure}[t]
\begin{center}
\includegraphics[width=\linewidth]{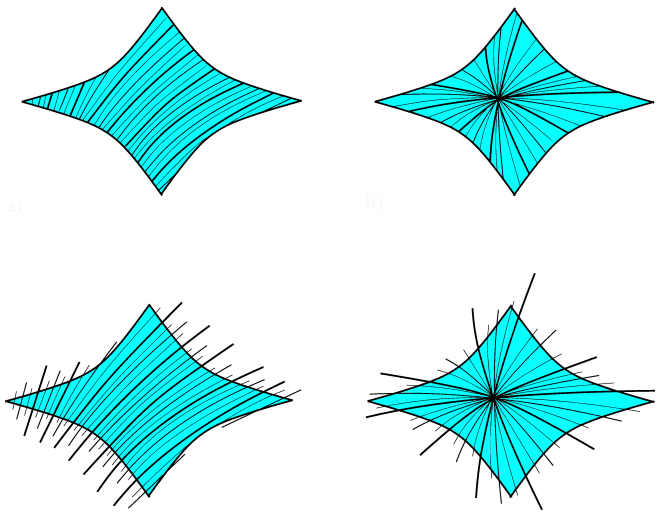}
\end{center}
\caption{The zones $\widehat{\Omega}_{\alpha}$ (top) and the zones 
$\Omega_{\alpha}$ (bottom) with adjoining segments corresponding to 
arising of unstable periodic trajectories on the Fermi surface 
(schematically, nets of directions of ${\bf B}$ inside the zones 
corresponding to stable periodic trajectories are also indicated).
}
\label{Fig6}
\end{figure}

 It can also be seen that when determining the zones $\Omega_{\alpha}$, 
as well as the Tsarev and Dynnikov directions for a fixed Fermi surface, 
one can use the functions $\tilde{\epsilon}_{1} ({\bf B})$ and 
$\tilde{\epsilon}_{2} ( {\bf B})$, while to determine the directions 
corresponding to arising of unstable periodic trajectories, it is 
necessary to know the functions $\epsilon_{1} ({\bf B} )$ and 
$\epsilon_{2} ({\bf B})$. The latter circumstance manifests itself, 
in particular, in a certain difference in the shape of zones $\Omega_{\alpha}$
from the zones $\widehat{\Omega}_{\alpha}$. Namely, the set of directions of 
${\bf B}$ corresponding to arising of open trajectories associated 
with the zone $\Omega_{\alpha}$ is somewhat larger than the zone itself 
and also contains an infinite set of segments adjacent to the boundary of
$\Omega_ {\alpha}$ and corresponding to arising of periodic 
trajectories on the Fermi surface (Fig. \ref{Fig6}). The periodic 
trajectories can then be considered stable for directions of ${\bf B}$ 
inside $\Omega_{\alpha}$ and unstable on additional segments.

  Such an arrangement of the zones $\Omega_{\alpha}$ actually 
leads to a rather complicated analytical behavior of the conductivity 
tensor near their boundaries, which makes it difficult to determine the 
shape of these zones in direct measurements of the conductivity 
(see, for example, \cite{AnalProp}). At the same time, however, there 
are methods for experimentally determining the exact mathematical 
boundaries of the zones $\Omega_{\alpha}$, which makes it possible 
to experimentally determine their exact shape (see \cite{ExactBound}).

 The zones $\Omega_{\alpha}$ represent regions with piecewise smooth 
boundaries on the sphere (see e.g. \cite{dynn3}). Other than that, 
we are not aware of any restrictions on their shape. In particular, there 
may be unconnected stability regions corresponding to the same values of 
$(m^{1}, m^{2}, m^{3})$. For simplicity, we agree here to consider the 
union of such domains as one disconnected stability zone on $\mathbb{S}^{2}$. 
In this sense, each region $\Omega_{\alpha}$ and its diametrically opposite 
one form the same stability zone. In addition, stability zones can also 
be non-simply connected (see e.g. \cite{DynMalNovUMN}). The latter, however, 
takes place for very specific Fermi surfaces, which are special mathematical 
examples. For real dispersion laws, we will assume here that all the zones 
$\Omega_{\alpha}$ are simply connected domains with piecewise smooth 
boundaries on $\mathbb{S}^{2}$.

 Below we give a brief description of various types of angular conductivity 
diagrams in metals, as well as their changes with a change in the value of 
$\epsilon_{F}$ in the interval $[ \epsilon_{\min} , \epsilon_{\max} ]$ 
(\cite{UltraCompl}), which we will need later. Here, we will be primarily 
interested in conductivity diagrams that correspond to dispersion laws with 
complex angular diagrams, i.e., diagrams containing an infinite number of 
zones $\widehat{\Omega}_{\alpha}$. 

\vspace{1mm}

  It is easy to see that if the value of $\epsilon_{F}$ is sufficiently 
close to the value $\epsilon_{\min}$ or $\epsilon_{\max}$, the Fermi 
surfaces are small ellipsoids, and open trajectories of system 
(\ref{MFSyst}) are absent on them for any direction of ${\bf B}$. 
It can also be noted that the Hall conductivity is of the electronic type 
in the first case and of the hole type in the second one. The corresponding 
conductivity diagrams can be called zero-type diagrams and denoted 
by $0_{-}$ or $0_{+}$, depending on the sign of the Hall conductivity.
In the general case, for a fixed dispersion relation $\epsilon ({\bf p})$, 
we can single out some values $\epsilon_{1}^{{\cal A}\prime}$ and 
$\epsilon_{2}^{{\cal A}\prime}$ such that the angular diagrams of the 
types $0_{-}$ and $0_{+}$ correspond to situations
$$\epsilon_{F} \, \in \, (\epsilon_{\min}, \epsilon_{1}^{{\cal A}\prime})
\quad \text{\rm and} \quad
\epsilon_{F} \, \in \, (\epsilon_{2}^{{\cal A}\prime}, \epsilon_{\max})$$
respectively.

 For generic dispersion relations, we can also single out the values 
$\epsilon_{1}^{\cal A}$ and $\epsilon_{2}^{\cal A}$, such that the situations
$$\epsilon_{F} \, \in \, (\epsilon_{1}^{{\cal A}\prime},
\epsilon_{1}^{\cal A})
\quad \text{\rm and} \quad
\epsilon_{F} \, \in \, (\epsilon_{2}^{\cal A}, 
\epsilon_{2}^{{\cal A}\prime})$$
correspond to conductivity diagrams containing only one-dimensional 
arcs corresponding to arising of unstable periodic trajectories 
on the Fermi surface. Diagrams of this type can be denoted by the symbols 
$1_{-}$ and $1_{+}$ depending on the type of the Hall conductivity observed 
for directions of ${\bf B}$ corresponding to the presence of only closed 
trajectories on the Fermi surface.

 The interval $(\epsilon_{1}^{\cal A}, \epsilon_{2}^{\cal A})$ corresponds 
to conductivity diagrams containing stability zones $\Omega_{\alpha}$. 
We can say that such diagrams have a sufficient level of complexity, and 
it is they that will be mainly of interest to us here. For generic dispersion 
relations with complex angular diagrams (with an infinite number of zones 
$\widehat{\Omega}_{\alpha}$), however, it is natural to divide this interval 
into three intervals (see \cite{UltraCompl})
$$\epsilon_{1}^{\cal A} \,\,\, < \,\,\, \epsilon_{1}^{\cal B} \,\,\, < \,\,\,
\epsilon_{2}^{\cal B} \,\,\, < \,\,\, \epsilon_{2}^{\cal A} $$

 Conductivity diagrams corresponding to the situations
$$\epsilon_{F} \, \in \, (\epsilon_{1}^{\cal A},
\epsilon_{1}^{\cal B})
\quad \text{\rm and} \quad
\epsilon_{F} \, \in \, (\epsilon_{2}^{\cal B}, 
\epsilon_{2}^{\cal A}) \,\,\, ,$$
can be called diagrams of the $A_{-}$ and $A_{+}$ types, respectively.
For diagrams of this type, in all regions of directions of ${\bf B}$ 
corresponding to the presence of only closed trajectories on the Fermi 
surface, the Hall conductivity has the same type (electronic and hole, 
respectively).

 Conductivity diagrams corresponding to the situation
$$\epsilon_{F} \, \in \, 
(\epsilon_{1}^{\cal B}, \epsilon_{2}^{\cal B}) \,\,\, , $$
can be called diagrams of type $B$. These diagrams are specified by the fact 
that in the space of directions of ${\bf B}$ (on the unit sphere $\mathbb{S}^{2}$), 
among the regions corresponding to the presence of only closed trajectories 
on the Fermi surface, there are both regions of the electronic Hall 
conductivity, and regions of the hole Hall conductivity.

 In fact, there are also two additional important differences between 
diagrams of type $A$ and diagrams of type $B$ (see \cite{UltraCompl}).

\vspace{1mm}

\noindent
1) Generic diagrams of type $A$ contain a finite number of zones 
$\Omega_{\alpha}$, while generic diagrams of type $B$ contain an infinite 
number of stability zones.

\vspace{1mm}

\noindent
2) Generic diagrams of type $A$ do not contain directions of ${\bf B}$ 
corresponding to arising of Tsarev or Dynnikov trajectories, while 
generic diagrams of type $B$ contain such directions.

\vspace{1mm}

\begin{figure*}[t]
\begin{center}
\includegraphics[width=\linewidth]{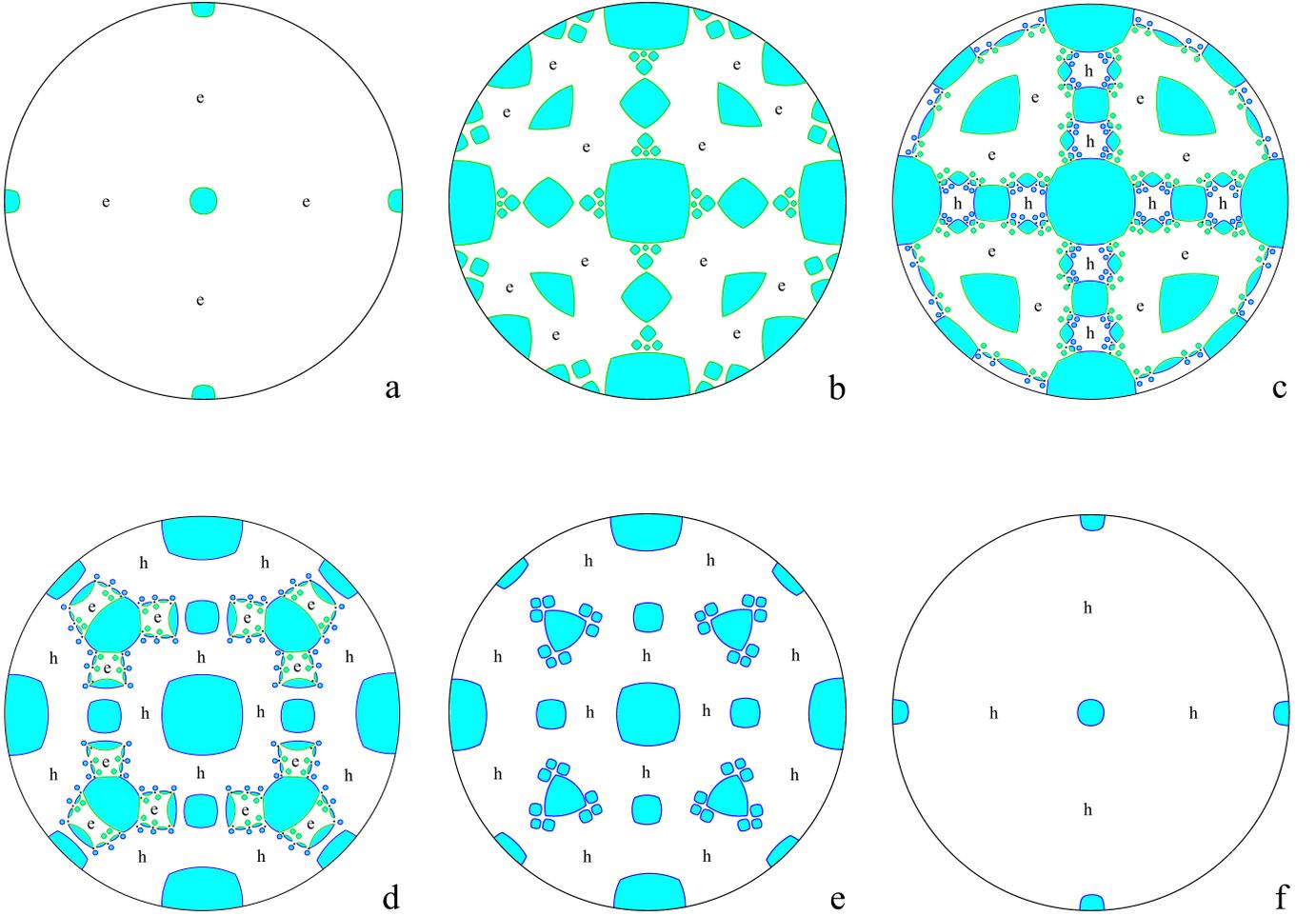}
\end{center}
\caption{Possible evolution of the conductivity diagram when 
$\epsilon_{F}$ changes from $\epsilon_{1}^{\cal A}$ to 
$\epsilon_{2}^{\cal A}$ and passes through $\epsilon_{ 1}^{\cal B}$ 
and $\epsilon_{2}^{\cal B}$. The zones $\Omega_{\alpha}$ and the 
electronic (e) and hole (h) Hall conductivity regions are shown 
schematically. We can distinguish diagrams of type $A_{-}$ ((a) and (b)),
diagrams of type $B$ ((c) and (d)) and diagrams of type $A_{+}$ ((e) and (f)).
}
\label{Fig7}
\end{figure*}

 Fig. \ref{Fig7} (schematically) shows a possible evolution of the
conductivity diagram in the situation we describe when $\epsilon_{F}$ 
changes from $\epsilon_{1}^{\cal A}$ to $\epsilon_{2}^{ \cal A}$.

\vspace{1mm}

 Diagrams in the interval $(\epsilon_{1}^{\cal B}, \epsilon_{2}^{\cal B})$  
contain also stability zones with a more complex boundary 
than in the intervals $(\epsilon_ {1}^{\cal A}, \epsilon_{1}^{\cal B})$ 
and $(\epsilon_{2}^{\cal B}, \epsilon_{2}^{\cal A}) $. Namely, here we have 
zones, part of the boundary of which is adjacent to the regions of the electronic 
Hall conductivity, and part to the regions of the hole Hall conductivity.

 We also note here that the parts of the boundaries of $\Omega_{\alpha}$ 
adjacent to the electronic Hall conductivity regions are determined by the 
relation $\tilde{\epsilon}_{1} ({\bf B}) = \epsilon_{F}$, and the parts of 
the boundaries adjacent to the hole Hall conduction regions are determined
by the relation $\tilde{\epsilon}_{2} ({\bf B}) = \epsilon_{F}$.

\vspace{1mm}

 The above picture corresponds to generic dispersion relations with angular 
diagrams containing an infinite number of zones $\widehat{\Omega}_{\alpha}$. 
This picture, in fact, may have the following degenerations.

1) $\epsilon_{1}^{{\cal A}\prime} = \epsilon_{1}^{\cal A}$ or
$\epsilon_{2}^{\cal A} = \epsilon_{2}^{{\cal A}\prime}$, such that the 
corresponding interval
$(\epsilon_{1}^{{\cal A}\prime}, \epsilon_{1}^{\cal A})$ or
$(\epsilon_{2}^{\cal A}, \epsilon_{2}^{{\cal A}\prime})$ shrinks to a point.
In the above picture, in this case, there are no diagrams of the type $1_{-}$ 
or $1_{+}$, such that diagrams of the type $0_{-}$ and $A_{-}$ or 
$A_{+}$ and $0_{+ }$ (or both) immediately pass into each other. Such 
degeneracies often arise for dispersion relations with sufficiently high 
symmetry (for example, cubic).

2) Degeneration $\epsilon_{1}^{\cal B} = \epsilon_{2}^{\cal B}$. 
In this case, there are no diagrams of type $B$ in the above picture,
and the diagram arising at the level 
$\epsilon_{1}^{\cal B} = \epsilon_{2}^{\cal B}$ coincides with the 
angular diagram for the entire dispersion relation. The corresponding 
dispersion relations form a very special class (of infinite codimension 
in the space of periodic $\epsilon ({\bf p})$) and we will not consider 
them here. We emphasize here only that in this case we have in mind 
dispersion relations whose diagrams remain complex (contain an infinite 
number of stability zones), despite the presence of degeneracy.
In addition to such cases, there are also deformations of the dispersion 
relations, under which the interval 
$(\epsilon_{1}^{\cal B}, \epsilon_{2}^{\cal B})$ is infinitely narrowed, 
and the corresponding angular diagrams are simplified and become 
diagrams with one stability zone at the degeneracy point. Such 
degenerations, in a sense, correspond to the boundary between the 
dispersion relations of the two types described above and are observed 
in a much more general situation.

\vspace{1mm}

 Although we are primarily interested here in dispersion relations 
with an infinite number of stability zones, let us also present here a typical 
picture of the change in the conductivity diagram for relations corresponding 
to the presence of only one zone $\widehat{\Omega}$. As before, we will 
assume here that all stability zones are simply connected, which corresponds 
to realistic dispersion relations that arise in real conductors.

 As in the previous case, for generic dispersion relations, here we can also 
introduce a set of reference points
$$\epsilon_{\min} \, < \, \hat{\epsilon}^{{\cal A}\prime}_{1} \, < \,
\hat{\epsilon}^{\cal A}_{1} \, < \, \hat{\epsilon}^{\cal B}_{1} \, < \,
\hat{\epsilon}^{\cal B}_{2} \, < \, \hat{\epsilon}^{\cal A}_{2} \, < \, 
\hat{\epsilon}^{{\cal A}\prime}_{2} \, < \, \epsilon_{\max} \,\,\, , $$
separating intervals with diagrams of different types.

 The intervals $[\epsilon_{\min}, \hat{\epsilon}^{{\cal A}\prime}_{1})$ and
$(\hat{\epsilon}^{{\cal A}\prime}_{2}, \epsilon_{\max}]$, and also
$(\hat{\epsilon}^{{\cal A}\prime}_{1}, \hat{\epsilon}^{\cal A}_{1})$ and
$(\hat{\epsilon}^{\cal A}_{2}, \hat{\epsilon}^{{\cal A}\prime}_{2})$,
as before, correspond here to diagrams of the types $0_{-}$, $0_{+}$, 
$1_{-}$, and $1_{+}$, respectively.

 The intervals $(\hat{\epsilon}^{\cal A}_{1}, \hat{\epsilon}^{\cal B}_{1})$ 
and $(\hat{\epsilon}^{\cal B}_{2}, \hat{\epsilon}^{\cal A}_{2})$ correspond 
to the diagrams $A_{-}$ and $A_{+}$ respectively. The only peculiarity here 
is that on such diagrams there is only one stability zone corresponding to 
a single set $(m^{1}, m^{2}, m^{3})$. The region that does not belong to the 
stability zone corresponds to the electronic Hall conductivity for diagrams 
of the type $A_{-}$, and to the hole conductivity for diagrams of the type 
$A_{+}$.

 The diagram appearing in the interval 
$(\hat{\epsilon}^{\cal B}_{1}, \hat{\epsilon}^{\cal B}_{2})$ will be called 
here a diagram of type $\widehat{B}$. It contains a single stability zone 
covering the entire unit sphere $\mathbb{S}^{2}$.

\vspace{1mm}

 As in the previous case, the above picture admits degenerations. 
In particular, the situations
$\, \hat{\epsilon}^{{\cal A}\prime}_{1} = \hat{\epsilon}^{\cal A}_{1} \, $
and
$\, \hat{\epsilon}^{\cal A}_{2} = \hat{\epsilon}^{{\cal A}\prime}_{2} \, $
correspond here to the same types of degeneracy as in the case of complex 
angular diagrams. The degeneration
$\, \hat{\epsilon}^{\cal B}_{1} = \hat{\epsilon}^{\cal B}_{2} \, $
corresponds to the ``boundary'' between dispersion relations with complex 
angular diagrams and those with simple angular diagrams.

\vspace{1mm}

\section{Lifshitz transitions and general principles of changing 
of conductivity diagrams}
\setcounter{equation}{0}

 Changing of the picture of open trajectories of system (\ref{MFSyst}) 
can be quite simple and visual for fairly simple Fermi surfaces. 
An illustrative example is the classical reconstruction considered 
in \cite{Lifshits1960} (Fig. \ref{Fig1}), where the compact Fermi surface 
takes the form of a warped cylinder. It is easy to see that open 
trajectories arise in this case only for directions of ${\bf B}$ 
orthogonal to the cylinder axis and are periodic. In the general case, 
however, the description of open trajectories on the Fermi surface 
is a rather complicated problem and often requires serious numerical 
studies (see e.g. \cite{DynnBuDA,DeLeo1,DeLeoObzor}). In this paper, 
we will try to describe the most general features of the changes in 
angular diagrams during the Lifshitz transitions, based on the general 
topological results obtained in the study of the Novikov problem. 
As we will see below, such features can lead in this case to a number 
of very special regimes in the conduction behavior, which are observed 
experimentally and are inherent precisely to situations close 
to topological transitions.

\begin{figure}[t]
\begin{center}
\includegraphics[width=\linewidth]{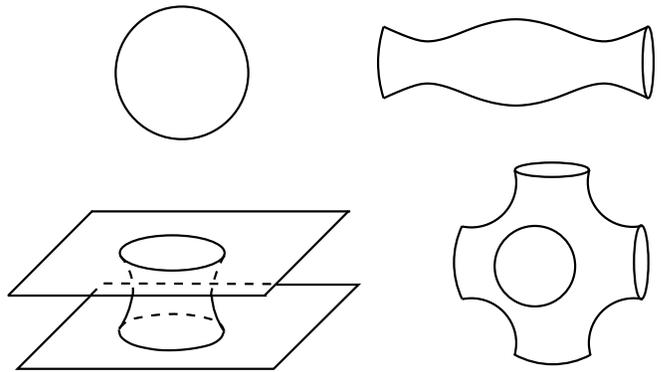}
\end{center}
\caption{Examples of Fermi surfaces of rank 0, 1, 2 and 3
}
\label{Fig8}
\end{figure}

 A natural indicator of the topological complexity of a Fermi surface 
is its rank, namely, the number of independent directions in which the 
surface extends in ${\bf p}$ - space. It is easy to see that the rank of 
the Fermi surface can take on the values 0, 1, 2 and 3 (Fig. \ref{Fig8}).

\begin{figure}[t]
\begin{center}
\includegraphics[width=\linewidth]{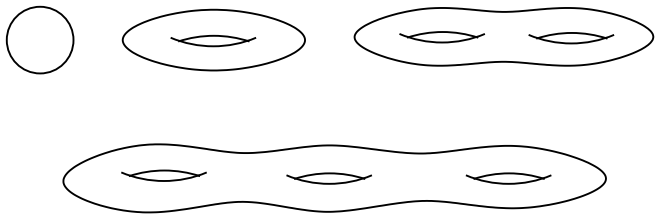}
\end{center}
\caption{Topological surfaces of genus 0, 1, 2 and 3
}
\label{Fig9}
\end{figure}

 Moreover, since the Fermi surface can also be considered as a compact 
surface in a three-dimensional torus, it also has topological genus $g$, 
which can take the values 0, 1, 2, 3, 4, ... and so on. (Fig. \ref{Fig9}). 
For topological reasons, the rank of a Fermi surface cannot exceed its 
genus. It is also important that, in addition to the topological complexity, 
the Fermi surface can have a very complex geometric shape in 
the ${\bf p}$ - space, which also has a significant effect on the shape of 
the trajectories of system (\ref{MFSyst}).

 The passage of singular points of the relation $\epsilon ({\bf p})$ with 
increasing energy $\epsilon_{F}$ changes the topology of the Fermi surface. 
It is easy to see that the passage of the minima and maxima of 
$\epsilon ({\bf p})$ leads to arising and disappearance of (small) 
components of the Fermi surface, while the passage of saddle singular points 
leads to the merging or decay of individual components, or to a change in 
their genus. If we talk about a reconstruction of a connected component 
of the Fermi surface, then passing a saddle singular point of index 1 
increases its genus by one, while passing a singular point of index 2 
decreases its genus by one. More generally, passing a singular point of 
index 1 can also lead to a merging of individual components, while passing 
a point of index 2 can lead to a splitting of one component into two.

 According to the Morse theory, a smooth function $\epsilon ({\bf p})$ on the 
torus $\mathbb{T}^{3}$ has at least three saddle singular points of both 
index 1 and index 2 (and, of course, at least one minimum and maximum). 
Quite often, however, the number of saddle singular points of 
$\epsilon ({\bf p})$ exceeds the lower estimates given, and the genus of 
the Fermi surface is 4 or more.

 Certainly, a change in the topology of the Fermi surface often gives 
obvious indications of a possibility of arising of non-closed 
trajectories of system (\ref{MFSyst}) on it. This is especially true for 
changes in the rank of the Fermi surface, as well as arising 
of periodic open trajectories. Usually, in these cases, the Fermi surfaces 
have a fairly simple shape and correspond to fairly simple angular 
conductivity diagrams.

 It will be of interest to us here to consider the situation when 
the Lifshitz transitions occur on fairly complex Fermi surfaces, which also 
correspond to fairly complex conductivity diagrams. Since the structure 
of such diagrams is formed mainly by the pattern of stability zones on them, 
it will be of interest to us, first of all, to trace the changes in 
this pattern during topological transitions.

\vspace{1mm}

 Above, we described the evolution of the picture of stability zones on a 
complex conductivity diagram, starting from the moment they appear on the 
diagram until they completely disappear (Fig. \ref{Fig7}). As the value 
of $\epsilon_{F}$ increases, the diagram changes monotonically, such that 
the region corresponding to the presence of only closed trajectories on the 
Fermi surface and the electron Hall conductivity decreases monotonically 
(until it disappears), and the analogous region corresponding to the hole Hall 
conductivity increases monotonically (since its emerging). In particular, 
sections of the boundaries of $\Omega_{\alpha}$ adjacent to the first 
region move outside the stability zones, and sections adjacent to the 
second region move inside the zones. As we said above, segments of the 
first type are defined by the relation 
$\tilde{\epsilon}_{1} ({\bf B}) = \epsilon_{F}$, 
and segments of the second type are determined by the relation 
$\tilde{\epsilon} _{2} ({\bf B}) = \epsilon_{F}$.

 In energy intervals that do not contain reconstructions of the Fermi 
surface, the evolution of the conductivity diagram is continuous. At the 
same time, as was pointed out in \cite{dynn3}, the functions 
$\tilde{\epsilon}_{1} ({\bf B})$ and $\tilde{\epsilon}_{2} ({\bf B})$ 
can be locally constant in some domains on the unit sphere. This 
phenomenon is associated precisely with topological reconstructions of the 
surface $S_{F}$, and the values of these functions on such ``plateaus'' 
coincide with the energies at which the corresponding reconstructions 
(the Lifshitz transitions) are observed. As can be seen, the picture of 
stability zones can in this case ``jump'' in some region of 
$\mathbb{S}^{2}$ corresponding to a ``plateau'' of the function 
$\tilde{\epsilon}_{1 } ({\bf B})$ or $\tilde{\epsilon}_{2} ({\bf B})$.
 
 We will try to consider here in most detail possible changes in the 
conductivity diagram during the Lifshitz transitions, including a description 
of the regimes of behavior of the tensor $\sigma^{kl} ({\bf B})$ 
corresponding to such changes. As we said above, we will start with 
the cases corresponding to arising or disappearance of stable 
open trajectories on the Fermi surface.

\begin{figure}[t]
\begin{center}
\includegraphics[width=0.8\linewidth]{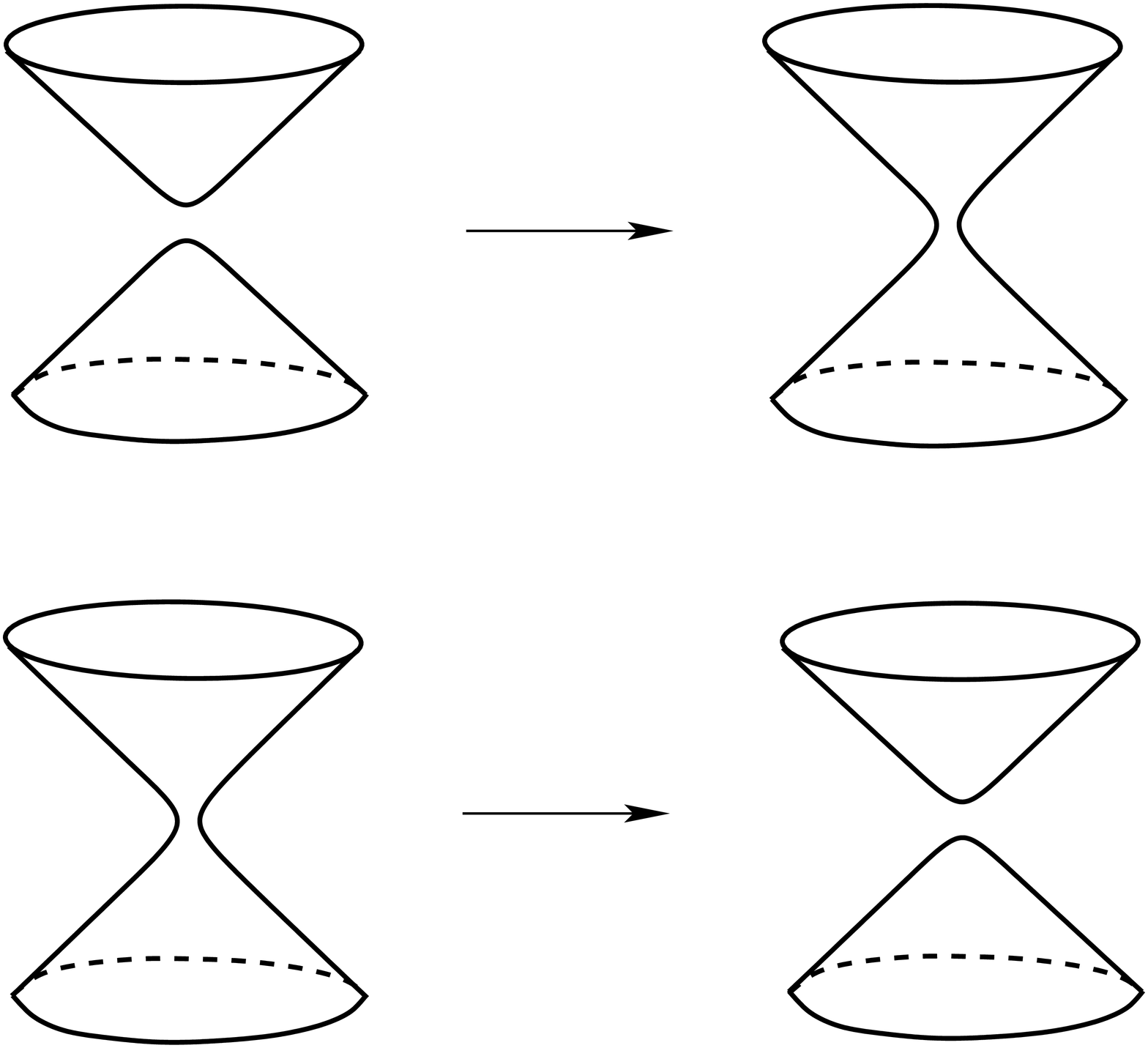}
\end{center}
\caption{Reconstructions of the Fermi surface when passing saddle 
singular points of $\epsilon ({\bf p})$ of index 1 (top) and 
index 2 (bottom)
}
\label{Fig10}
\end{figure}

 In connection with the study of open trajectories of system (\ref{MFSyst}), 
we will be interested in the structure of the Lifshitz transitions associated 
with the passage of saddle singular points of the relation $\epsilon ({\bf p})$. 
Fig. \ref{Fig10} shows the reconstructions of the Fermi surface when passing 
singular points of index 1 and 2 with increasing Fermi energy. 
Reconstructions in Fig. \ref{Fig10} look like mutually inverse, 
we must remember, however, that in both cases, 
for increasing $\epsilon_{F}$, the region 
$\, \epsilon ({\bf p}) < \epsilon_{F} \, $ increases and the region 
$\, \epsilon ({\bf p}) > \epsilon_{F} \, $ decreases.

\begin{figure}[t]
\begin{center}
\includegraphics[width=0.9\linewidth]{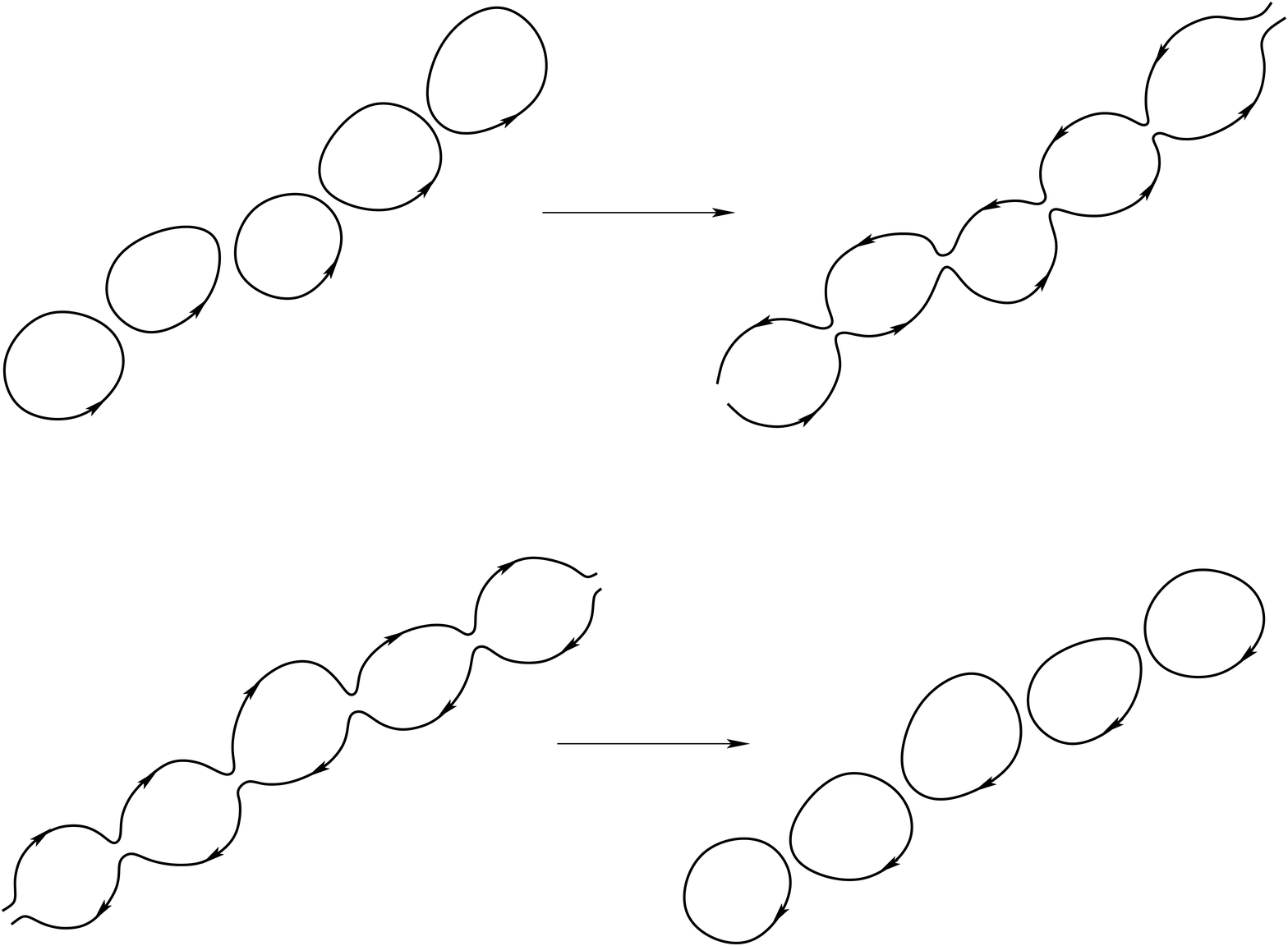}
\end{center}
\caption{Formation of open trajectories and their decay 
with increasing value of $\epsilon_{F}$
}
\label{Fig11}
\end{figure}

 All changes in the picture of open trajectories on the Fermi surface 
with increasing $\epsilon_{F}$ can be associated with two processes 
in planes orthogonal to ${\bf B}$, namely, the formation of open 
trajectories from closed electron-type trajectories and the decay of 
open trajectories into closed hole-type trajectories (Fig. \ref{Fig11}). 
Similarly, as the value of $\epsilon_{F}$ decreases, these processes 
go in the opposite direction.

\begin{figure}[t]
\begin{center}
\includegraphics[width=0.9\linewidth]{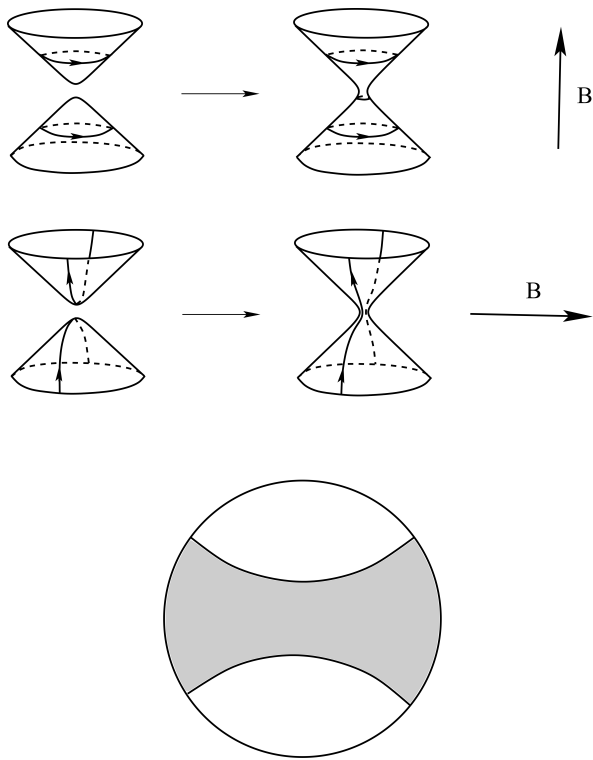}
\end{center}
\caption{Changes in trajectories when passing through a singular point 
for different directions of ${\bf B}$ and the area of possible changes 
in the conductivity diagram (shaded)
}
\label{Fig12}
\end{figure}

 It is easy to see that for directions of ${\bf B}$ close to the axis of 
the cone
\begin{equation}
\label{Konus1}
a^{2} d p_{1}^{2} + b^{2} d p_{2}^{2} - c^{2} d p_{3}^{2} \, = \, 0 
\end{equation}
(in the coordinate system corresponding to the given saddle singular point) 
or, respectively,
\begin{equation}
\label{Konus2}
a^{2} d p_{1}^{2} - b^{2} d p_{2}^{2} - c^{2} d p_{3}^{2} \, = \, 0 
\end{equation}
no changes in the picture of open trajectories of system 
(\ref{MFSyst}) can occur. The equation (\ref{Konus1}) or (\ref{Konus2}) 
thus selects two ellipsoidal regions on the unit sphere, in which 
the conductivity diagram (in our sense) certainly does not change when 
passing through the corresponding singular point. A change in the picture 
of open trajectories on the Fermi surface can thus occur only in the 
circular region separating opposite ellipsoidal regions on $\mathbb{S}^{2}$ 
(Fig. \ref{Fig12}). It can also be noted that in the case of observing 
of sharp changes along the boundary of this region (or part of it), it is 
not difficult to determine the parameters of the corresponding singular 
point (more precisely, the quantities $b/a$ and $c/a$).

 For further consideration, we need a brief description of the structure 
of system (\ref{MFSyst}) on the Fermi surface in the presence of stable 
open trajectories on it (see \cite{zorich1,dynn1,dynn3}). We give here 
this description using a model Fermi surface.

\begin{figure}[t]
\begin{center}
\includegraphics[width=\linewidth]{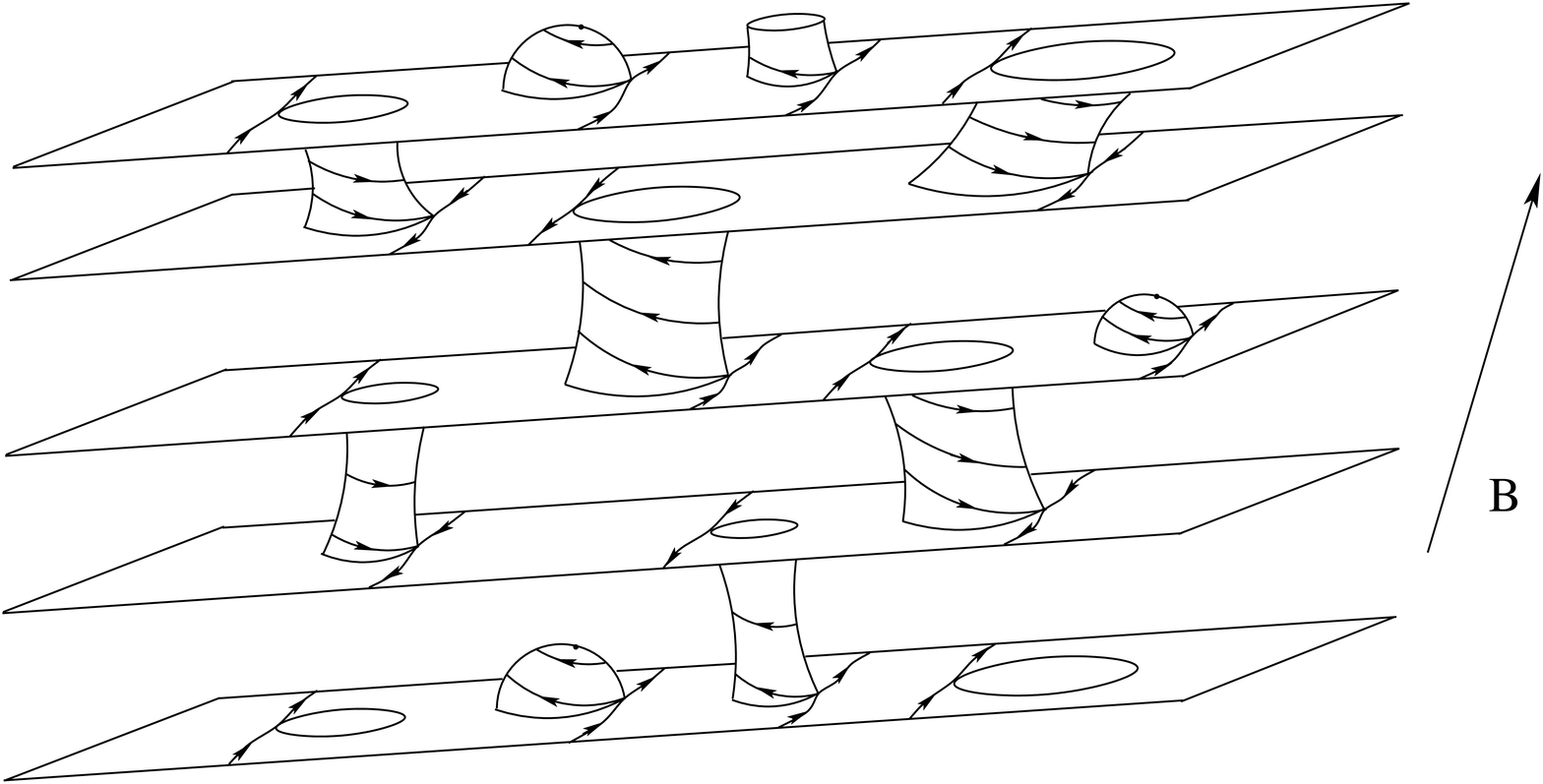}
\end{center}
\caption{Model Fermi surface carrying stable open trajectories 
of system (\ref{MFSyst})
}
\label{Fig13}
\end{figure}

 Consider in ${\bf p}$ - space a periodic family of integral planes 
connected by cylinders (Fig. \ref{Fig13}). As before, we call a plane in 
the ${\bf p}$ - space integral if it is generated by two reciprocal lattice 
vectors.

 We assume that the surface under consideration is periodic with periods 
equal to the reciprocal lattice vectors. In addition, we assume that all 
planes are divided into even and odd ones, so that the even planes remain 
even, and the odd planes remain odd, when shifted by any period.

\begin{figure}[t]
\begin{center}
\includegraphics[width=\linewidth]{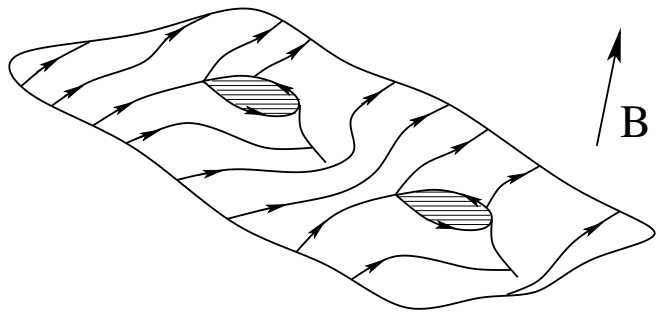}
\end{center}
\caption{Carrier of stable open trajectories of system (\ref{MFSyst}) 
on the Fermi surface
}
\label{Fig14}
\end{figure}

 It is easy to see that for directions of ${\bf B}$ almost orthogonal to 
the direction of the planes, our cylinders contain closed trajectories, 
separating the planes from each other (Fig. \ref{Fig13}). In this case, 
our planes contain stable open trajectories of system (\ref{MFSyst})  
with the mean direction given by the intersection of the plane orthogonal 
to ${\bf B}$ and the integral direction of the planes. The carriers 
of open trajectories are in this case (periodically deformed) planes with 
holes formed after the removing of closed trajectories (Fig. \ref{Fig14}).
It is also easy to see that the directions of open trajectories are opposite 
to each other on even and odd planes.

 The above picture is stable with respect to small rotations of ${\bf B}$ 
and is preserved as long as there are closed trajectories separating integral
planes on the cylinders connecting these planes. The corresponding stability 
zone $\Omega$ is obviously the larger, the higher and narrower the cylinders 
connecting the planes, and vice versa, is small for wide cylinders of low height.
It can also be noted that the disappearance of a cylinder of closed 
electron-type trajectories corresponds to sections of the boundary of 
$\Omega$ adjacent to the hole Hall conductivity regions, while the 
disappearance of a cylinder of closed hole-type trajectories corresponds 
to sections of the boundary adjacent to the regions of the electronic Hall 
conductivity.

 The presented picture is topological and geometrically it can look much 
more complicated. In particular, carriers of open trajectories can be deformed 
much more strongly, and the cylinders connecting them can have a very small 
height and a rather complex shape. Nevertheless, the described topological 
representation of the Fermi surface always arises when it contains stable 
open trajectories of system (\ref{MFSyst}) (\cite{zorich1,dynn1,dynn3}). 
This representation is not unique for a given Fermi surface; in particular, 
different such representations for the same surface arise in different 
stability zones.

\vspace{1mm}

 Let us now describe the possible changes in the picture of stability 
zones when passing a saddle point of index 1 (see Fig. \ref{Fig10}), 
using the above structure. Let us first consider the case when the 
reconstruction of the Fermi surface leads to the formation of stable 
open trajectories (Fig. \ref{Fig11}) for some direction of ${\bf B}$.
Using the structure described above, we will show here that the 
stability zones $\Omega_{\alpha}$ arising as a result of such 
a reconstruction have, in a certain sense, a special shape, and also 
a specific contribution to the conductivity tensor in the 
limit $\, B \rightarrow \infty \, $.

\begin{figure}[t]
\begin{center}
\includegraphics[width=0.8\linewidth]{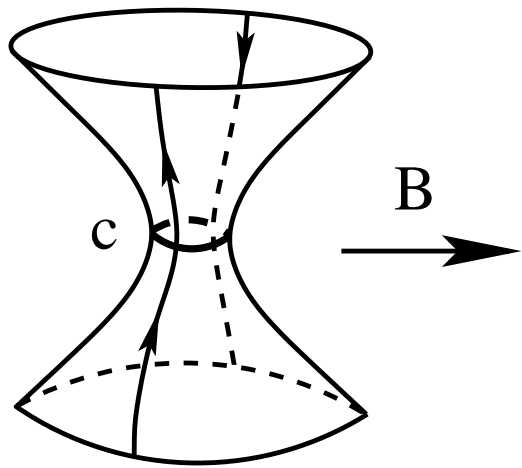}
\end{center}
\caption{A neck formed after a topological reconstruction 
and a cycle intersecting the resulting open trajectories
}
\label{Fig15}
\end{figure}

\begin{figure}[t]
\begin{center}
\includegraphics[width=0.8\linewidth]{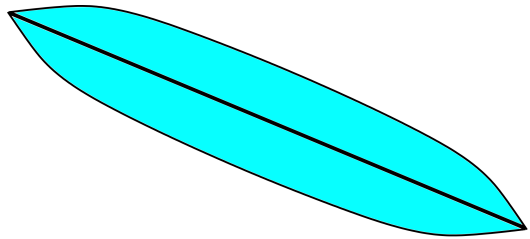}
\end{center}
\caption{A stability zone formed as a result of passing 
a singular point of index 1 with an increase in the value 
of $\epsilon_{F}$ (schematically)
}
\label{Fig16}
\end{figure}

 Since the arising of open trajectories occurs due to the 
reconstruction of the Fermi surface, all such trajectories must pass 
through a narrow neck that appears after the passage of the saddle 
singular point (Fig. \ref{Fig15}). This means, in particular, that 
the cycle $c$ shown in Fig. \ref{Fig15}, must pass both through 
the carrier of open trajectories running in one direction and 
through the carrier of open trajectories running in the opposite 
direction. From this it follows then that it also passes from one 
base of a cylinder of closed trajectories separating these carriers 
to its other base. It can be seen, therefore, that the height of 
at least one cylinder of closed trajectories connecting two carriers 
of open trajectories is very small and tends to zero when approaching 
the topological transition point. In addition, this cylinder has 
a saddle singular point at each of its bases, which are adjacent 
to two different carriers of open trajectories. It is not dificult 
to show then that such points must lie on different necks in 
the ${\bf p}$ - space, and the cylinder itself, thus, passes 
through both these necks. It is easy to see then that quite small 
rotations of ${\bf B}$, except for those orthogonal to the vector 
connecting the indicated necks, will lead to the disappearance of 
the indicated cylinder of closed trajectories and, thus, to going beyond 
the zone $\Omega_{\alpha}$. It can be seen, therefore, that the stability 
zone formed as a result of the reconstruction must be a very narrow 
region on the sphere $\mathbb{S}^{2}$ (Fig. \ref{Fig16}).

 As the transition point is approached, the width of the region 
$\Omega_{\alpha}$ tends to zero, so that $\Omega_{\alpha}$ tends 
to a one-dimensional arc on the unit sphere (Fig. \ref{Fig16}).
It is easy to see that this arc is a segment of a great circle 
orthogonal to some integer direction in the ${\bf p}$ - space 
(namely, to the vector connecting the two necks considered above).
At the points of this segment, therefore, the plane orthogonal 
to ${\bf B}$ always contains some fixed reciprocal lattice vector.
It is also not difficult to show that the width of the region 
$\Omega_{\alpha}$ tends to zero according to the law 
$\, \sim \sqrt{(\epsilon_{F} - \epsilon_{0})/\epsilon_{F}} \, $ 
when approaching the transition point.

 The passage of a saddle point of index 1 can lead to arising
of both a finite and an infinite number of narrow stability zones 
on the angular diagram. In the latter case, as can be seen, we should 
expect the arising of an angular diagram of type $B$ described 
above. 

  It can also be noted that the total area of the stability zones 
arising as a result of the topological reconstruction tends to zero 
when approaching the transition point. As a consequence, the passage 
of a singular point of index 1 does not cause an abrupt decrease in 
the area of the regions corresponding to the Hall conductivity of the 
electronic type (and the presence of only closed trajectories on the 
Fermi surface).

 The formation of stability zones when passing through a singular 
point of index 1 can occur on diagrams of the $0_{-}$, $1_{-}$, $A_{-}$, 
and $B$ types. In the very first case, as is easy to see, this leads 
to a change of the type $0_{-}$ immediately to the type $A_{-}$. 
As we have already said, such degeneracies are typical for dispersion 
relations of sufficiently high symmetry and arising of several 
singular points of the same type at one energy level.

\vspace{1mm}

\begin{figure}[t]
\begin{center}
\includegraphics[width=0.9\linewidth]{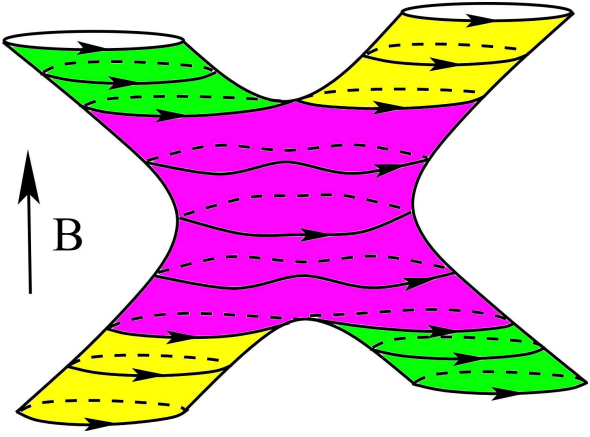}
\end{center}
\caption{An example of a Fermi surface consisting of cylinders of 
closed trajectories of the system (\ref{MFSyst})
}
\label{Fig17}
\end{figure}

 To describe the behavior of the conductivity tensor in our situation, 
we also need to discuss the measure of open trajectories that appear 
on the Fermi surface. In fact, for generic directions of ${\bf B}$ 
(maximal irrationality), the measure of such trajectories is small 
in the parameter of proximity to the transition point. To show this, 
consider the Fermi surface immediately before passing through 
(one or several) singular points of index 1. For generic directions 
of ${\bf B}$, there are only closed trajectories on it in this case, 
and the Fermi surface itself represents a set of a finite number of 
(non-equivalent) cylinders of closed trajectories separated by singular 
trajectories (Fig. \ref{Fig17}).

\begin{figure}[t]
\begin{center}
\includegraphics[width=\linewidth]{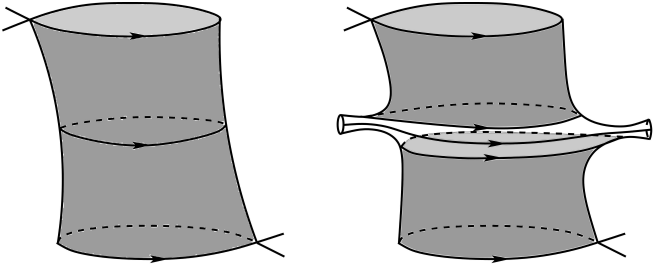}
\end{center}
\caption{Thin necks adjacent to a cylinder of closed trajectories 
of finite height and reconstruction of trajectories on the Fermi 
surface (cylinders of conserved trajectories are shaded)
}
\label{Fig18}
\end{figure}

 Near the transition point at $\, \epsilon_{F} > \epsilon_{0} \, $, 
thin necks appear on the Fermi surface, connecting its various parts. 
Considering such necks on each of the cylinders of closed trajectories, 
we can see that with a strong decrease in their diameter, most of the 
closed trajectories do not undergo any changes (Fig. \ref{Fig18}).
As a result, in the limit $\, \epsilon_{F} \rightarrow \epsilon_{0} \, $
almost all trajectories for such directions of ${\bf B}$ remain closed. 
It is also easy to see that the fraction of trajectories changed 
on each of the cylinders is proportional to the ratio of the neck 
width to the height of the cylinder, i.e. 
$\sqrt{(\epsilon_{F} - \epsilon_{0})/\epsilon_{F}}$.
The corresponding factor also arises for the contribution 
(\ref{Periodic}) of open trajectories to the conductivity tensor 
in the limit $\, \omega_{B} \tau \rightarrow \infty \, $ 
(as well as a factor containing a weak logarithmic dependence on 
$(\epsilon_{F} - \epsilon_{0})/\epsilon_{F}$ due to the proximity 
of the trajectories to singular points of system (\ref{MFSyst}) in 
the narrow necks).

 The above discussion, however, needs one important addition. Namely, 
in addition to the proximity to a topological transition point, in 
the situation under consideration, the proximity of the direction of 
${\bf B}$ to the directions corresponding to arising of periodic 
open trajectories on the Fermi surface (even before the transition 
point) can also play an important role. Such trajectories can exist 
on both sides of the topological transition and occupy a finite 
area on the Fermi surface. For directions of ${\bf B}$ close to such 
directions, at least some of the cylinders of closed trajectories 
described above (Fig. \ref{Fig17}) have small heights and 
large ``transverse'' sizes in ${\bf p}$ - space. In this case, 
the ratio of the neck diameter to the cylinder height can remain finite.

\begin{figure}[t]
\begin{center}
\includegraphics[width=\linewidth]{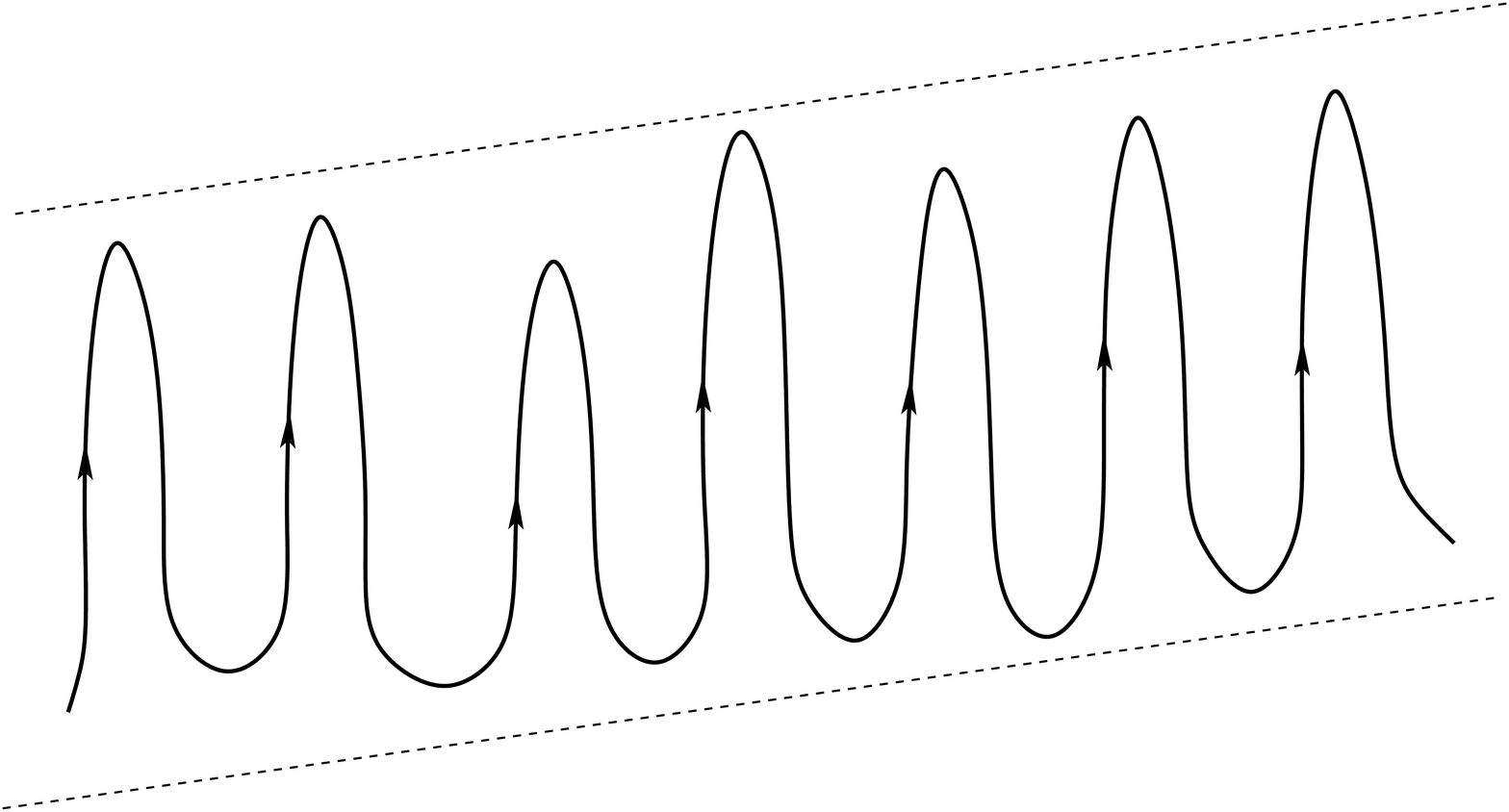}
\end{center}
\caption{Stable open trajectories arising for directions of 
${\bf B}$ close to directions of arising of periodic trajectories 
(schematically) 
}
\label{Fig19}
\end{figure}

 The proximity of generic directions of ${\bf B}$ to the directions 
described above may be caused by the specific geometry of the stability 
zones (Fig. \ref{Fig16}). This applies, first of all, to the limit 
segment inside the zone which can be a set of directions corresponding 
to arising of periodic trajectories even before the transition 
point. As we have seen above, the width of the region $\Omega_{\alpha}$ 
is also proportional to $\sqrt{(\epsilon_{F} - \epsilon_{0})/\epsilon_{F}}$ 
and the same we can also say about the height of a part of the cylinders 
of closed trajectories that arise for our directions of  
$\, {\bf B} \in \Omega_{\alpha} \, $ before the topological transition 
(when there are no open trajectories for these directions yet).
As a consequence, the measure of open trajectories on the Fermi surface 
in the zone $\Omega_{\alpha}$ can remain finite near the transition. 
In this case, however, the emerging stable open trajectories have 
a specific geometry. Namely, they are limited by straight strips of 
rather large width and repeat the geometry of periodic trajectories 
on small scales (see Fig. \ref{Fig19}).

 For the described directions of ${\bf B}$, we can introduce a 
function $\mu ({\bf B})$, which determines the ratio of the minimal 
width of strips containing open trajectories to the size of the 
Brillouin zone. The contribution of open trajectories to 
the conductivity at $\, \omega_{B} \tau \gg 1 \, $ is different 
in the intervals $\, 1 \ll \omega_{B} \tau \leq \mu ({\bf B} ) \, $ 
and $\, \omega_{B} \tau \gg \mu ({\bf B}) \, $. In the first case, 
this contribution can be approximated by the formula (\ref{Periodic}), 
provided that the direction of the $x$ axis coincides with the 
direction of periodic open trajectories. In the second case, 
the direction of the $x$ axis must coincide with the mean direction 
of stable open trajectories in ${\bf p}$ - space, and the total 
contribution of such trajectories to the conductivity 
tensor can be represented as
\begin{equation}
\label{muForm}
\sigma^{kl} \,\,\, \simeq \,\,\,
{n e^{2} \tau \over m^{*}}  \left(
\begin{array}{ccc}
\mu^{2} ( \omega_{B} \tau )^{-2}  &  ( \omega_{B} \tau )^{-1}  &
\mu ( \omega_{B} \tau )^{-1}   \\
( \omega_{B} \tau )^{-1}  &  \mu^{-2}  &  \mu^{-1}   \\
\mu ( \omega_{B} \tau )^{-1}  & \mu^{-1}  &  *
\end{array}  \right)  
\end{equation} 
($\omega_{B} \tau \, \rightarrow \, \infty$).

 The function $\mu ({\bf B})$ has a strong dependence on the direction 
of ${\bf B}$ and goes to infinity for directions corresponding to the 
arising of periodic trajectories that exist on both sides of the 
transition.

\vspace{1mm}

 The zones $\Omega_{\alpha}$, which have additional (rotational) 
symmetry and appear as a result of emerging of several singular 
points of index 1 at the same level $\epsilon_{0}$, also deserve 
special mention. The sizes of such zones tend to zero in all directions 
as the transition point is approached, and the measure of open 
trajectories on the Fermi surface at 
$\, {\bf B} \in \Omega_{\alpha} \, $ is proportional to 
$\sqrt{(\epsilon_ {F} - \epsilon_{0})/\epsilon_{F}}$.
 
\begin{figure}[t]
\begin{center}
\includegraphics[width=\linewidth]{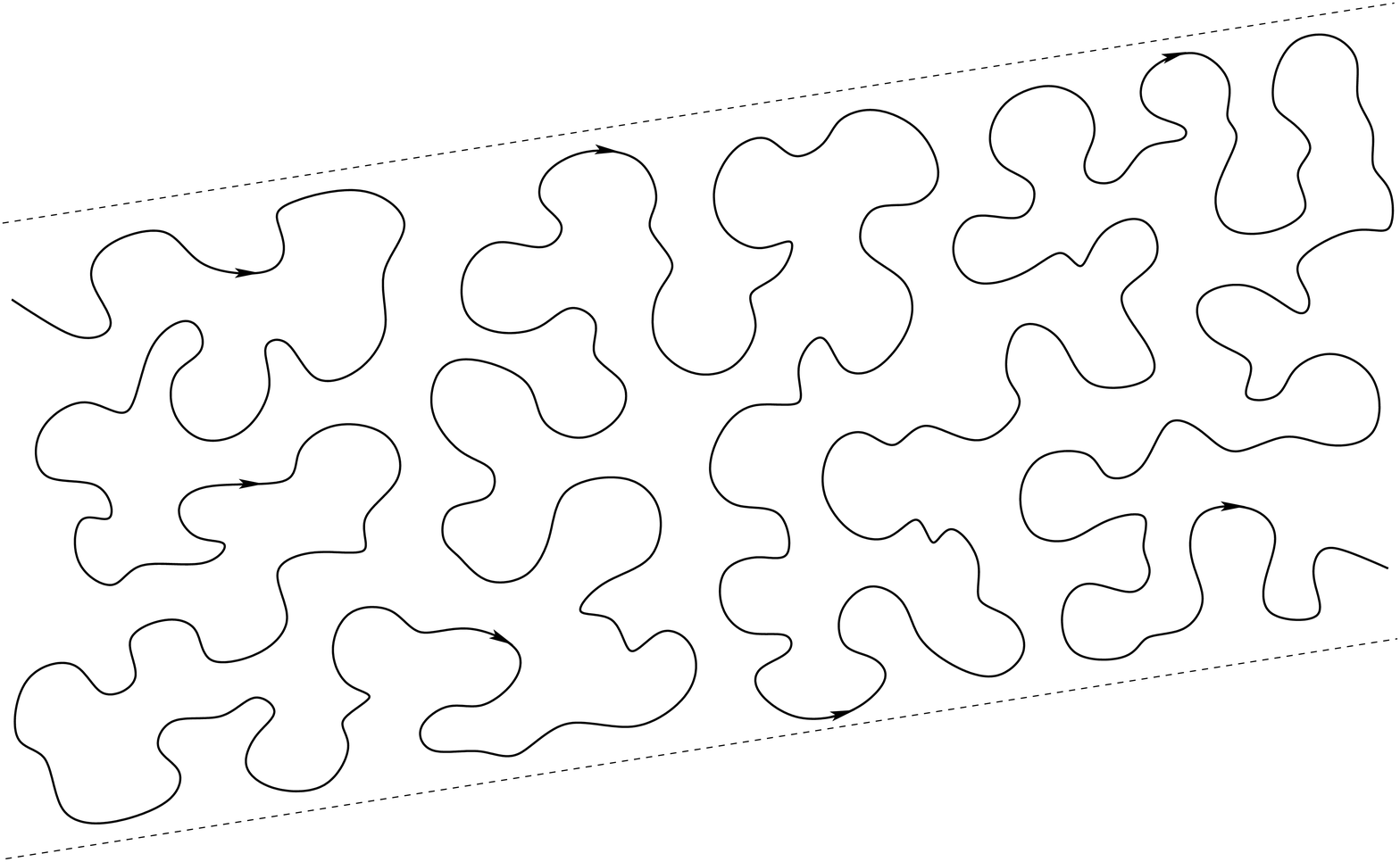}
\end{center}
\caption{Stable open trajectories arising for directions of 
${\bf B}$ lying in a symmetric stability zone of small sizes
}
\label{Fig20}
\end{figure}

 Symmetric stability zones, however, have one more peculiarity.
Namely, the vector $(m^{1}, m^{2}, m^{3})$ for such zones coincides 
with the direction passing through the center of the zone 
(see \cite{dynn3}), and for this direction of ${\bf B}$ there are 
no open trajectories on the Fermi surface. For directions of 
${\bf B}$ lying in a symmetric zone of small sizes, stable open 
trajectories lie in straight strips of large width and have rather 
complex behavior on small scales (see Fig. \ref{Fig20}).

 To describe the contribution to the conductivity given by open 
trajectories in symmetric stability zones arising during 
the Lifshitz transitions, it is also natural to consider the 
intervals $\, 1 \ll \omega_{B} \tau \leq \mu ({\bf B}) \, $ 
and $ \, \omega_{B} \tau \gg \mu ({\bf B}) \, $ for the 
function $\mu ({\bf B})$, which has the same meaning as above. 
In the first interval, the behavior of the conductivity is 
more complicated (the arising of intermediate fractional 
powers of $\omega_{B} \tau$ is possible), while the total 
contribution of such trajectories to the conductivity tensor 
also contains a small factor of the order of 
$\sqrt{(\epsilon_{F} - \epsilon_{0})/\epsilon_{F}}$.
In the second interval, the contribution of open trajectories 
to the conductivity is similar to the contribution 
(\ref{muForm}), and also multiplied by a factor of the order of 
$\sqrt{(\epsilon_{F} - \epsilon_{0})/\epsilon_{F}}$.
It is likely that the observation also of a weak logarithmic 
dependence on $(\epsilon_{F} - \epsilon_{0})/\epsilon_{F}$, 
due to the proximity to singular points of $\epsilon ({\bf p})$, 
is somewhat complicated here from the experimental point of 
view.

\vspace{1mm}

 It must be said that the conductivity in the region of the 
small stability zones, arising as a result of 
the Lifshitz transitions, as a whole, has a rather complex behavior, 
and it is probably more convenient to study the geometry of such 
zones using methods that differ from methods of direct 
measurements of conductivity (see, for example, \cite{ExactBound}). 
At the same time, the arising of such zones plays a very important 
role in changing the structure of a general conductivity diagram, 
especially in the case of arising of complex diagrams of 
the type $B$.

\vspace{1mm}

 Let us now consider the second possible situation, namely, the 
disappearance of stable open trajectories of system (\ref{MFSyst}) 
when passing through a singular point of index 1 (Fig. \ref{Fig11}). 
In this situation, therefore, we will talk about the disappearance 
of a stability zone or a part of it. The corresponding changes, 
obviously, can occur only on diagrams of the types $A_{-}$, $B$, 
and $A_{+}$. 

 As we have already seen, the decay of stable open trajectories of 
system (\ref{MFSyst}) becomes possible when the carriers of such 
trajectories are no longer separated from each other and the 
possibility of ``jumping'' between such carriers appears. Thus, 
a topological reconstruction of the Fermi surface leads to the 
decay of stable open trajectories if, as a result of the 
reconstruction, a new cylinder is formed that connects two 
carriers and makes it possible to ``jump'' between them for 
a given direction of ${\bf B}$. This is exactly the situation 
that leads to the formation of plateaus in the values of the 
functions $\tilde{\epsilon}_{1} ({\bf B})$ and 
$\tilde{\epsilon}_{2} ({\bf B} )$ (in our case, of 
$\tilde{\epsilon}_{2} ({\bf B})$) (see \cite{dynn3}), 
which, in turn, should lead to jumps in the conductivity 
diagrams, considered by us here.

 As we have already noted, a change in the conductivity diagram 
during our reconstructions can occur only in a special circular 
region (Fig. \ref{Fig12}). With a change of considered type, 
we can observe an instantaneous disappearance of zones of 
finite sizes (or their parts) at the moment of the topological 
transition. In addition, the measure of open trajectories on the 
Fermi surface here also remains finite up to the moment of 
transition and instantly vanishes (for generic directions 
of ${\bf B}$) when the surface is reconstructed. 

\begin{figure}[t]
\begin{center}
\includegraphics[width=\linewidth]{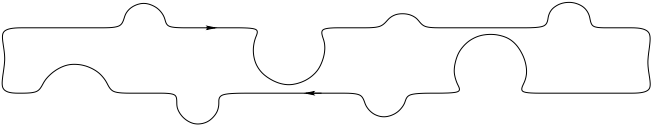}
\end{center}
\caption{Long closed trajectories arising on the Fermi surface 
near a topological transition (schematically)
}
\label{Fig21}
\end{figure}

 At the same time, the proximity to the Lifshitz transition 
affects the specifics of the trajectories of system (\ref{MFSyst}), 
as well as the behavior of the conductivity tensor, also in the 
described situation. In this case, it is expressed in  
arising of very long closed trajectories of system 
(\ref{MFSyst}) on the Fermi surface for generic directions of 
${\bf B}$ lying in the disappeared stability zones or their 
disappeared parts (see Fig. \ref {Fig21}). The average length 
of such trajectories tends to infinity near the transition and 
decreases away from it. In addition, as we noted above, the 
values of the functions $\epsilon_{1} ({\bf B})$ and 
$\epsilon_{2} ({\bf B})$ differ from the values of 
$\tilde{\epsilon} _{1} ({\bf B})$ and 
$\tilde{\epsilon}_{2} ({\bf B})$ for directions of ${\bf B}$ 
corresponding to arising of periodic open trajectories.
As a consequence of this, the periodic trajectories of 
system (\ref{MFSyst}) on the Fermi surface do not disappear 
immediately at the moment of transition, but persist for 
some time. As a result, the region of a vanished stability 
zone (or its vanished part) is covered by a net of 
one-dimensional arcs corresponding to the presence of 
periodic trajectories on the Fermi surface. The net of 
corresponding arcs becomes denser when approaching 
a topological transition and becomes infinitely dense 
at the moment of transition.

 The described features of the trajectories of (\ref{MFSyst}) 
near the Lifshitz transition also lead to a rather complicated 
behavior of the conductivity tensor in strong magnetic fields. 
The corresponding behavior of $\sigma^{kl} ({\bf B})$ was 
described in \cite{AnalProp}, where it appeared in very narrow 
regions near the boundaries of the zones $\Omega_{\alpha}$ 
in the conductivity diagrams. Here, however, such behavior 
occurs in rather large areas, namely, in the place of the 
disappeared zones $\Omega_{\alpha}$ or their parts. In such 
domains, for generic directions of ${\bf B}$, it is natural 
to introduce an (approximate) function $\lambda ({\bf B})$, 
which determines the ratio of the average size of long closed 
trajectories to the size of the Brillouin zone. In addition, 
when considering the conductivity in these regions, it is 
natural to keep the coordinate system corresponding to the 
disappeared open trajectories, namely, choosing the mean
direction of the former open trajectories in 
the ${\bf p}$ - space as the $x$ axis.

 A natural consequence of the geometry of trajectories 
in the regions under consideration is that their contribution 
to the conductivity manifests itself as the contribution of 
closed trajectories under the much stronger condition 
$\, \omega_{B} \tau \gg \lambda ({\bf B}) \, $, while in the 
interval $\, 1 \ll \omega_{B} \tau \leq \lambda ({\bf B}) \, $ 
their contribution rather corresponds to that of open trajectories.
However, even in the limit 
$\, \omega_{B} \tau \gg \lambda ({\bf B}) \, $, the contribution 
of the resulting closed trajectories preserves anisotropy in the 
plane orthogonal to ${\bf B}$.

 In addition, it can be seen that the long closed trajectories are 
formed from open trajectories located on two different carriers.
For dispersion relations satisfying the condition 
$\, \epsilon ({\bf p}) = \epsilon (-{\bf p}) \, $ 
(and Fermi surfaces of not too large genus), this actually 
implies the relation 
$\, \langle v^{z}_{gr} \rangle \rightarrow 0 \, $ at 
$\, \lambda ({\bf B}) \rightarrow \infty \, $ for the 
trajectory-averaged electron group velocity along the 
direction of magnetic field. As a consequence, the contribution 
of such trajectories to the conductivity along the magnetic field 
actually tends to zero in the limit 
$\, \omega_{B} \tau \gg \lambda ({\bf B}) \gg 1 \, $. 
In the latter, such a contribution is similar to the 
contribution to the conductivity given by unstable Dynnikov's 
trajectories in the limit $\, \omega_{B} \tau \gg 1 \, $.

 In general, the total contribution of the long closed trajectories 
to the symmetric $s^{kl}$ and antisymmetric $a^{kl}$ parts of 
the conductivity tensor in the limit 
$\, \omega_{B} \tau \gg \lambda ({\bf B }) \, $ can be represented 
as (\cite{AnalProp})
\begin{multline*}
s^{kl} (B) \,\,\, \simeq \,\,\,   \left(
\begin{array}{ccc}
0 \,\,\,\,\, &  0  &  \,\, 0   \cr
0  \,\,\,\,\, &  0  &  \,\, 0   \cr
0  \,\,\,\,\, &  0  &  \sigma^{zz} (\lambda) 
\end{array}
\right)  \,\,\, +   \\
+ \,\,\, {n e^{2} \tau \over m^{*}} \left(
\begin{array}{ccc}
(\omega_{B} \tau)^{-2}  &  \lambda (\omega_{B} \tau)^{-2}  &
\lambda (\omega_{B} \tau)^{-2}   \cr
\lambda (\omega_{B} \tau)^{-2}  &
\lambda^{2} (\omega_{B} \tau)^{-2}  &
\lambda^{2} (\omega_{B} \tau)^{-2}  \cr
\lambda (\omega_{B} \tau)^{-2}  &
\lambda^{2} (\omega_{B} \tau)^{-2}  &
\lambda^{2} (\omega_{B} \tau)^{-2}
\end{array}
\right)  
\end{multline*}
(where $\, \sigma^{zz} (\lambda) \rightarrow 0 \, $ at 
$\lambda \rightarrow \infty $),
$$a^{kl} (B) \,\,\, \simeq \,\,\,
{n e^{2} \tau \over m^{*}} \left(
\begin{array}{ccc}
0  &  (\omega_{B} \tau )^{-1}  &  (\omega_{B} \tau )^{-1}  \cr
(\omega_{B} \tau )^{-1}  &  0  &  (\omega_{B} \tau )^{-1}  \cr
(\omega_{B} \tau )^{-1}  &  (\omega_{B} \tau )^{-1}  &  0
\end{array}
\right) $$

 It must be said that the condition 
$\, \omega_{B} \tau \gg \lambda ({\bf B}) \, $ can be quite strong, 
and in many cases some intermediate regime between the regime 
(\ref{Periodic}) and the dependence described above can be observed. 
We also note that, in the general case, in addition to the contribution 
described above, we must also add the contribution (\ref{Closed}) 
from ``ordinary'' closed trajectories, which are also present on the 
Fermi surface in the described situation.

 The value $\lambda ({\bf B})$ goes to infinity at the 
topological transition point. For a fixed generic direction of 
${\bf B}$, its behavior near $\epsilon_{0}$ can be (approximately) 
described by the dependence
$$\lambda \, \sim \, \sqrt{\epsilon_{F}/|\epsilon_{F} - \epsilon_{0}|} $$

 At the same time, the dependence of $\lambda$ on the direction of 
${\bf B}$ is quite complicated, in particular, $\lambda$ goes to 
infinity on the (preserved) arcs corresponding to arising
of periodic trajectories of (\ref{MFSyst}). In general, vanishing 
stability zones (or parts of them) can be called regions of complex 
conductivity behavior in strong magnetic fields.

\vspace{1mm}

\begin{figure*}[t]
\begin{center}
\includegraphics[width=\linewidth]{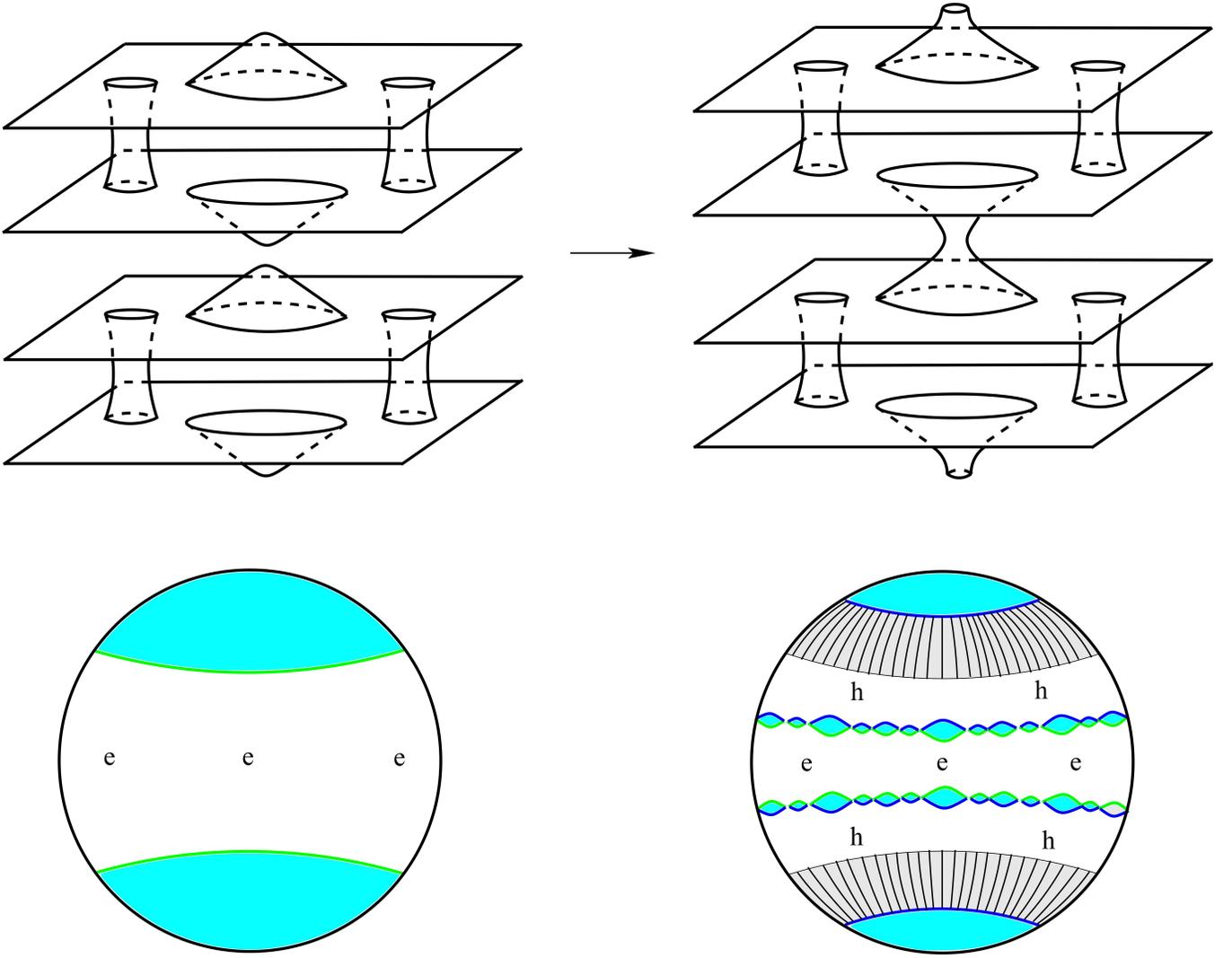}
\end{center}
\caption{An example of a passage of a saddle singular point of index 1 
and the corresponding change in the angular conductivity diagram 
(schematically, the change in the picture of stability zones and the 
formation of a region of complex conductivity behavior are shown)
}
\label{Fig22}
\end{figure*}

 In fact, both described effects (arising and disappearance of 
stability zones) can be observed simultaneously (in different parts 
of an angular diagram) when passing through a saddle singular point 
of index 1. Fig. \ref{Fig22} shows an example of one of such reconstructions 
of the Fermi surface. It is easy to see that before the reconstruction the 
angular diagram is rather simple (of the type $A_{-}$) and contains 
one stability zone (with a diametrically opposite one). After 
the reconstruction, a part of the stability zone disappears being 
replaced by a zone of complex conductivity behavior. In addition, 
many small zones arise that separate the region of the electron Hall 
conductivity from the region of the hole Hall conduction that appears 
in the diagram. It can be shown that, in the immediate vicinity of the 
topological transition, chains of small stability zones are located 
very close to zones of complex conductivity behavior, while further away 
from the transition they shift towards the ``equator''. In general, the 
conductivity diagram after the transition is of type $B$ and, in the generic 
case, must also contain directions of ${\bf B}$ corresponding to arising
of unstable trajectories of the Tsarev or Dynnikov type.

\vspace{1mm}

  We note here that the passage of singular points with an 
increase in the value of $\epsilon_{F}$ can cause abrupt changes 
in the conductivity diagrams, however, preserving the general 
direction of their evolution, shown in Fig. \ref{Fig7}. 
As a consequence, the disappearance of a stability zone or part 
of it leads in this situation to an abrupt increase in the area 
corresponding to the hole Hall conductivity (and presence of only 
closed trajectories on the Fermi surface). At the same time, the 
area of the region corresponding to the electronic Hall conductivity 
(and presence of only closed trajectories on the Fermi surface) 
immediately near the transition remains unchanged. It can be seen, 
therefore, that if the described effect (disappearance of stable 
open trajectories when passing through a singular point of index 1) 
takes place for a diagram of type $A_{-}$, the type of the diagram 
changes to type $B$. From the same considerations, we can conclude 
that the observation of the described effect on diagrams of 
type $B$ does not change the type of a diagram. Observation of 
the described effect on a diagram of type $A_{+}$ preserves 
the type of the diagram or changes it to type $1_{+}$.

 In connection with what has been said above, we would like to 
note here another important circumstance. Namely, diagrams of 
type $B$ are generic diagrams and should, generally speaking, 
be observed in the study of sufficiently rich families of Fermi 
surfaces of sufficiently complex shape. However, they have not 
yet been experimentally discovered. The same is also true for 
unstable trajectories of the Tsarev or Dynnikov type, which 
must accompany such diagrams. One of the reasons for this, 
in our opinion, may be a rather small width of the interval 
$(\epsilon^{\cal B}_{1}, \epsilon^{\cal B}_{2})$ and, accordingly, 
a low probability of the value $\epsilon_{F}$ falling into it 
for real dispersion relations. In this regard, it can be expected 
that the use of the Lifshitz transitions can provide good opportunities 
for observing such diagrams, as well as nontrivial regimes of 
conductivity behavior corresponding to arising of Tsarev 
or Dynnikov trajectories on the Fermi surface.
 
 In general, the effects we have discussed can produce changes in 
the conductivity diagrams listed below, with the following possible 
changes in the diagram type 
$$\begin{array}{c}
0_{-} \rightarrow A_{-}  \\ 
1_{-} \rightarrow A_{-}  \\ 
A_{-} \rightarrow \{ A_{-}, B \}  \\ 
B \rightarrow B  \\ 
A_{+} \rightarrow \{ 1_{+}, A_{+} \}  
\end{array} $$

\vspace{1mm}

 The effects described above correspond to the passage of a saddle 
singular point of index 1 in the ``forward'' direction, namely, as 
the Fermi energy $\epsilon_{F}$ increases. Certainly, as a result 
of an external influence, the Lifshitz transitions can be performed both 
in the ``forward'' and ``backward'' directions. Moreover, an external 
action does not necessarily simply change the Fermi level, but 
generally changes the entire dispersion relation $\epsilon ({\bf p})$.
Obviously, in the general case, the topology of a reconstruction of 
the Fermi surface is actually determined by changes in the relation 
$\epsilon ({\bf p})$ near the corresponding singular point.
It is natural, however, to keep the terms ``the passage of 
a singular point in the forward or backward direction'', based on the 
coincidence of the topology of the corresponding transition with the 
topology of the transition with increasing or decreasing $\epsilon_{F}$.

 Considering the change in the relation $\epsilon ({\bf p})$ to be 
continuous, in a rather narrow neighborhood of a topological 
transition, the influence of the general change in 
$\epsilon ({\bf p})$ can be neglected and it can be assumed that 
the main changes in the picture of stability zones are caused 
precisely by the reconstruction of the Fermi surface. In order to 
describe the corresponding changes in the angular diagram, we can use 
the picture obtained in the consideration given above. As we have 
already said, the above picture refers to the passage of a singular 
point of index 1 in the forward direction. In the general case, 
we are interested here in a similar description for passing a
point of index 1 in the backward direction, as well as passing 
a point of index 2 in the forward and backward directions.

\vspace{1mm}

 The passage of a singular point of index 1 in the backward direction 
naturally leads to effects opposite to those described above. Namely, 
changing of stable open trajectories when passing a saddle point of 
index 1 in the backward direction can produce changes in the conductivity 
diagrams of the types listed below, with the following possible changes 
in the type of diagram:
$$\begin{array}{c}
A_{-} \rightarrow \{ 0_{-}, 1_{-},  A_{-} \}  \\
B \rightarrow  \{ A_{-}, B \} \\ 
A_{+} \rightarrow A_{+}  \\
1_{+} \rightarrow A_{+}
\end{array} $$

 In this case, the picture of stability zones in the conductivity 
diagram can undergo the following specific changes

\vspace{1mm}

1) Reducing the size of a certain (finite or infinite) number of 
stability zones and their disappearance directly at the topological 
transition point.

\vspace{1mm}

2) Formation of regions of complex conductivity behavior of finite 
size in the region of the hole Hall conductivity when approaching a 
topological transition and their transformation into stability zones 
at the transition point (or their attachment to already existing zones).

\vspace{1mm}

 The passage of a saddle singular point of index 2 in the forward 
direction is in fact similar to the passage of a singular point of 
index 1 in the backward direction, however, with the replacement of 
the electronic Hall conductivity regions by the hole Hall conductivity 
regions, and vice versa. Thus, a change in stable open trajectories 
when passing a saddle singular point of index 2 in the forward 
direction can produce changes in the conductivity diagrams of 
the types listed below with the following possible changes in 
the diagram type:
$$\begin{array}{c}
1_{-} \rightarrow A_{-}  \\ 
A_{-} \rightarrow A_{-}  \\ 
B \rightarrow  \{ A_{+}, B \} \\ 
A_{+} \rightarrow \{ 0_{+}, 1_{+},  A_{+} \} 
\end{array} $$

 In this case, the picture of stability zones in the conductivity 
diagram can undergo the following specific changes.

\vspace{1mm}

1) Reducing the size of a certain (finite or infinite) number of 
stability zones and their disappearance directly at the topological 
transition point.

\vspace{1mm}

2) Formation of regions of complex conductivity behavior of finite 
size in the region of the electronic Hall conductivity when approaching 
a topological transition and their transformation into stability 
zones at the transition point (or their attachment to already 
existing zones).

\vspace{1mm}

 Similarly, changing of stable open trajectories when passing a saddle 
singular point of index 2 in the backward direction can produce changes 
in the conductivity diagrams of the types listed below, with the following 
possible changes to the diagram type:
$$\begin{array}{c}
A_{-} \rightarrow \{ 1_{-}, A_{-} \}  \\
B \rightarrow B  \\
A_{+} \rightarrow \{ A_{+}, B \}  \\
1_{+} \rightarrow A_{+}  \\
0_{+} \rightarrow A_{+}  
\end{array} $$

 In this case, the picture of stability zones in the conductivity 
diagram can undergo the following specific changes.

\vspace{1mm}

1) Arising (of a finite or infinite number) of new stability 
zones of zero size at the transition point and a gradual increase 
in their size with further distance from it.

\vspace{1mm}

2) Disappearance of some stability zones of finite size 
(or their parts) with the formation of regions of complex conductivity 
behavior with the electronic type of Hall conductivity. If this effect 
is observed on a $A_{+}$ diagram, it turns into a $B$ type diagram, 
with the arising of an infinite number of stability zones, 
as well as directions of ${\bf B}$ corresponding to the occurrence 
of Tsarev's or Dynnikov's trajectories on the Fermi surface.

\vspace{1mm}

 It can also be noted here that each of the above descriptions can also 
be used in the case of passing several singular points of the same type 
at once. Such a situation, in fact, can arise quite often for dispersion 
relations with additional (rotational) symmetry.

 As for the simultaneous passage of singular points of different types, 
such a situation is a nongeneric situation and is observed only for 
special dispersion relations. In particular, it may refer to relations 
separating relations with simple angular diagrams (with a single 
stability zone) and relations with complex angular diagrams.

 Each of the above descriptions of the changes in conductivity diagrams, 
certainly, can also be used for dispersion relations with simple angular 
diagrams, taking into account the peculiarities of the evolution of 
conductivity diagrams for such relations.

\vspace{1mm}

 In conclusion, we would like to note that although the above picture 
refers primarily to changes in the structure of stability zones in 
conductivity diagrams, it also, in fact, describes many features of 
arising and disappearance of unstable open trajectories of 
system (\ref{MFSyst}) on the Fermi surface under the described 
reconstructions. Moreover, if we talk about trajectories of the 
Tsarev or Dynnikov type, their arising is uniquely related with 
diagrams of the type $B$ and, thus, is completely determined by the 
picture of stability zones on the unit sphere.

 If we consider unstable periodic open trajectories of system 
(\ref{MFSyst}), then, as can be seen, most of them are also associated 
with stability zones and appear either near their edges or in regions 
of complex conductivity behavior. As can also be seen, most of these 
trajectories do not change directly at the topological transition point, 
but undergo changes at some distance from it, following the changes in the 
corresponding stability zone.

 Among the unstable periodic trajectories of system (\ref{MFSyst}), 
however, we should also especially note the trajectories that are not 
tied to any specific stability zone on the conductivity diagram. 
In fact, the corresponding directions of ${\bf B}$ almost always 
belong to some zones $\widehat{\Omega}_{\alpha}$ for the entire 
dispersion relation, and the corresponding trajectories are consistent 
with stable open trajectories, occurring in these zones. However, when 
they appear far from the corresponding interval 
$(\tilde{\epsilon}_{1} ({\bf B}) , \tilde{\epsilon}_{2} ({\bf B}))$, 
they usually are not very interesting from this point of view.
Instead, however, they are usually closely related to the topology 
of a given Fermi surface and exhibit its geometric properties well. 
A representative example of such trajectories are the periodic trajectories 
considered in \cite{Lifshits1960}. It is easy to see that changes 
associated with the arising or disappearance of such trajectories 
correspond to transitions between diagrams of type 0 and 1, 
or do not change the diagram type.

\section{Conclusion}
\setcounter{equation}{0}

 The paper considers the topological Lifshitz transitions in metals 
and related changes in galvanomagnetic phenomena from the point 
of view of the general Novikov problem. Namely, the picture of 
possible changes in electron trajectories in a metal during 
topological reconstructions of the Fermi surface and the 
corresponding changes in the behavior of electrical conductivity 
in the presence of strong magnetic fields is considered.
The consideration is based on the classification of non-closed 
electron trajectories arising on Fermi surfaces of arbitrary 
complexity, and of corresponding behavior of conductivity  
in strong magnetic fields. The main analysis is based on the 
description of possible changes in the picture of stable open 
trajectories on the Fermi surface during topological transitions.
As shown in the paper, the Lifshitz transitions are accompanied by 
a certain number of such changes, which make it possible to 
determine the features of the transition topology based on the 
observation of the conductivity in strong magnetic fields.
The results obtained in the work can serve as one of the tools 
for studying topological Lifshitz transitions in metals with 
complex Fermi surfaces.

\vspace{5mm}

 The study was supported by the grant of the Russian Science 
Foundation \textnumero $\,\, $ 21-11-00331, ``Geometric methods 
in the Hamiltonian theory of integrable and almost integrable systems''

\end{document}